\documentclass[letterpaper,dvipsnames]{article}
\pdfoutput=1
\usepackage{jheppub}
\usepackage{amsmath}
\usepackage{graphicx}
\usepackage{array}
\usepackage{ulem}
\usepackage{cancel}
\usepackage{xcolor}

\usepackage{slashed}
\usepackage{verbatim}
\usepackage{afterpage}
\usepackage{appendix}
\usepackage{multirow}
\usepackage{float}
\usepackage{mathtools}
\mathtoolsset{centercolon}

\usepackage{multirow}

\newcommand{\EM}{{\rm em}}

\newcommand{\tr}{\text{tr}}

\newcommand{\red}{\textcolor{red}}

\title{Testing F-theory GUTs with the Axiverse}

\author[a]{Michael Nee,}
\affiliation[a]{Department of Physics, Harvard University, Cambridge, MA, 02138, USA}
\emailAdd{mnee@fas.harvard.edu}

\author[b]{Mario Reig}
\affiliation[b]{Theoretical Physics Department, CERN, 1211 Geneva 23, Switzerland}
\emailAdd{mario.reig.lopez@cern.ch}

\author[c,d]{and Timo Weigand}
\affiliation[c]{II. Institut f\"ur Theoretische Physik, Universit\"at Hamburg, Notkestrasse 9, 22607 Hamburg, Germany}
\affiliation[d]{Zentrum f\"ur Mathematische Physik, Universit\"at Hamburg, Bundesstrasse 55, 20146 Hamburg, Germany}
\emailAdd{timo.weigand@desy.de}

\abstract{We show that axions coupled to photons in F-theory Grand Unified Theories (GUTs) satisfy the coupling-to-mass relation $g_{a\gamma}/m_a \leq C\, \frac{\alpha_{\rm em}}{2\pi}\frac{1}{m_\pi f_\pi}$, with $C$ a calculable coefficient. 
This bound, which is saturated for the QCD axion with $C = \mathcal{O}(1)$, has previously been shown to hold in field theoretic and perturbative heterotic GUT constructions.
 In F-theory, topological GUT symmetry breaking by hypercharge flux introduces axion-like particles (ALPs) coupled to photons without coupling to QCD. These ALPs arise from the non-universal holomorphic threshold corrections to the gauge kinetic functions induced by the hypercharge flux. 
 When gauge couplings approximately unify near the string scale, as required in phenomenologically viable models, the shift symmetries of these ALPs are broken by D-instantons whose action is controlled by the size of the threshold corrections to the gauge couplings. Small corrections imply unsuppressed instantons and heavy ALPs. We compute the resulting axion potentials and show that the coupling-to-mass ratio $g_{a\gamma}/m_a$ for every non-universal ALP lies well below the QCD axion prediction. 
   We consider possible loopholes to this result -- some of which could lead to $C\gg 1$ -- and argue that none of them allows for $g_{a\gamma}/m_a$ to be arbitrarily above the QCD axion prediction within regions of control for the effective action.
 The bound persists in models with large threshold corrections, where new incomplete GUT multiplets at intermediate energy scales are required. 
   As a result, in the geometric regime, where the $\alpha'$ expansion is under control, no ALP parametrically above the QCD axion band exists. Our results make F-theory GUTs falsifiable: finding an ALP far above the QCD band, for example discovering axion-induced cosmic birefringence,  rules out 
   F-theory GUTs (along with field theoretic and perturbative heterotic GUT constructions) in regimes of control of the effective theory. Finally, we briefly discuss the F-theory GUT expectations for the QCD axion mass as well as the impact of heavy ALPs in cosmology.}

\begin{document}

\preprint{CERN-TH-2026-129}

\maketitle

\section{Introduction}

The hint for gauge coupling unification at energies close to the 4-dimensional Planck scale together with the peculiar Standard Model (SM) quantum numbers are two of the strongest arguments in favour of Grand Unified Theories (GUTs)~\cite{Georgi:1974sy}. Despite this success, minimal 4d GUTs present phenomenological challenges such as incorrect relations between SM fermion masses and the so-called doublet-triplet splitting problem. The latter calls for a dynamical mechanism explaining why the SM Higgs doublet remains light and the colour triplet scalar receives a large mass avoiding the constraints from proton decay searches~\cite{Super-Kamiokande:2020wjk}.

Some of the shortcomings of minimal GUTs are more easily addressed in theories where unification occurs in higher dimensions. Solutions to these problems can be obtained in bottom-up approaches such as the so-called orbifold GUTs~\cite{Kawamura:1999nj,Kawamura:2000ev,Altarelli:2001qj,Hall:2001pg,Hebecker:2001wq} and top-down constructions like the heterotic string~\cite{Gross:1984dd}. In the latter, for example, one can employ non-simply connected manifolds where GUT symmetry breaking can occur by turning on discrete Wilson lines~\cite{Candelas:1985en,Witten:1985xc,Green:1987mn}. These constructions allow for light Higgs doublets without the dangerous colour triplet scalars and, at the same time, avoid the incorrect fermion mass predictions. Theories based on the $E_8\times E_8$ heterotic string nicely reproduce standard GUT phenomenology but other aspects of string model building, such as moduli stabilisation (see~\cite{McAllister:2023vgy} for a recent review), appear more challenging in heterotic models compared to D-brane constructions, which allow a local approach to string model building. Weakly coupled type II string constructions using D-branes, as reviewed for instance in \cite{Blumenhagen:2006ci,Ibanez:2012zz}, are, however, harder to reconcile with standard GUT phenomenology, as recently summarised in \cite{Marchesano:2022qbx,Marchesano:2024gul}. 

This picture changed with the realisation that F-theory \cite{Vafa:1996xn}, a non-perturbative formulation of the strong coupling limit of the type IIB string, offers new approaches to GUT model building \cite{Donagi:2008ca,Beasley:2008dc,Beasley:2008kw, Donagi:2008kj}. This framework nicely combines some of the desirable properties of heterotic and D-brane model building, allowing one to construct GUT realisations of the SM in local models which need not know about the global structure of the internal manifold. For example, while $SO(10)$ spinors are absent in weakly coupled D-brane models, they can be obtained in theories with 7-branes if the string coupling becomes large enough. Similar strong coupling phenomena allow for exceptional gauge symmetries. In addition to Higgsing and discrete Wilson lines, F-theory constructions admit a new topological mechanism to break the GUT symmetry: turning on a line bundle and associating its structure group to the SM hypercharge group $U(1)_Y$~\cite{Beasley:2008kw,Donagi:2008kj}. 
 As a result, the GUT group can be broken down to the SM without inducing a St\"uckelberg mass for the hypercharge gauge boson that remains massless \cite{Buican:2006sn}. For more background and a guide to the literature we also refer to reviews such as \cite{Heckman:2010bq,Weigand:2010wm,Weigand:2018rez,Marchesano:2022qbx,Marchesano:2024gul}.

Testing the framework of GUTs, either in their 4d field theoretic form or in more stringy versions, has proven to be notoriously difficult. The most renowned prediction of these theories is proton decay with rates that parametrically scale as $\Gamma_{\rm proton} \sim m_{\rm proton}^5/M_{\rm GUT}^4$. This process has not been found in Super-Kamiokande~\cite{Super-Kamiokande:2020wjk}, which currently constrains $M_{\rm GUT}\gtrsim  10^{16}$ GeV. This bound to the GUT scale is expected to be improved by a factor of $\sim 2$ at Hyper-Kamiokande~\cite{Abe:2011ts} in a near future. Despite these promising prospects, it is crucial to search for complementary laboratory and cosmological tests of unification. This need for alternative probes is reinforced by the fact that different mechanisms, from imposing symmetries on top of minimal GUT models to simply enhancing the value of $M_{\rm GUT}$ by an $\mathcal{O}(1)$ factor, could make the decay of protons unobservable even at Hyper-Kamiokande.

Recently, it has been shown that due to its topological properties the axion-photon coupling is sensitive to unification of the SM gauge group~\cite{Agrawal:2022lsp,Agrawal:2025rbr,Agrawal:2024ejr,Reig:2025dqb}. The axion-gauge boson coupling is unaffected by renormalization, so this result is independent of the symmetry breaking pattern or the value of the GUT scale, $M_{\rm GUT}$.
In these theories, it is possible to show that the coupling-to-mass ratio for any axion  -- be it the QCD axion or a more generic axion-like particle (ALP) -- satisfies the relation
\begin{equation}
    \label{eq:coupling-to-mass-ratio}\frac{g_{a\gamma}}{m_a} \leq C\frac{\alpha_{\rm em }}{2\pi}\frac{1}{m_\pi f_\pi} \,,
\end{equation}
with $C=(E/N-1.92)$ being a model dependent $\mathcal{O}(1)$ coefficient that depends on the EM and QCD anomaly coefficients, $E$ and $N$, as well as the mixing of the QCD axion with the neutral mesons. This result comes about as any axion that couples to photons in a GUT must also couple to QCD, thus generating a mass for the axion. Measuring the photon coupling-to-mass ratio, $g_{a\gamma}/m_a$, offers the possibility of testing 4d GUTs~\cite{Agrawal:2022lsp}, orbifold GUTs~\cite{Agrawal:2025rbr}, as well as the weakly coupled heterotic string~\cite{Agrawal:2024ejr,Reig:2025dqb} in the laboratory~\cite{Kahn:2016aff,Melcon:2018dba,Ouellet:2018beu,Marsh:2018dlj,Ouellet:2019tlz,Lawson:2019brd,Beurthey:2020yuq,Schutte-Engel:2021bqm,Salemi:2021gck,DMRadio:2022pkf,DMRadio:2022jfv,Aja:2022csb,Bourhill:2022alm,ALPHA:2022rxj,Benabou:2022qpv,Berlin:2019ahk,Giaccone:2022pke,DMRadio:2023igr,Oshima:2023csb,DeMiguel:2023nmz,Ahyoune:2023gfw,Alesini:2023qed,BREAD:2023xhc,CAST:2024eil,Kalia:2024eml,Friel:2024shg,Baryakhtar:2025jwh,Ankel:2026zrv,Esposito:2026dhq}, astrophysics~\cite{Caputo:2023cpv,Regis:2024znx,EPTA:2024gxu,Benabou:2024jlj,Ning:2024eky,Ning:2025fqd,DeMarchi:2026pak,TerolCalvo:2026nlr}  and in cosmology~\cite{Minami:2020odp,Eskilt:2022cff,Diego-Palazuelos:2023mpy,Galaverni:2023zhv,Diego-Palazuelos:2025dmh,Carralot:2026kps}.\footnote{Finding a fractionally charged particle at colliders or other experiments would also inform us about the existence or not of 4d GUTs, see \cite{Alonso:2024pmq,Koren:2024xof,Koren:2025utp} for recent studies.}

Not only are axions generic in theories of quantum gravity~\cite{Svrcek:2006yi,Arvanitaki:2009fg}, including type IIA~\cite{Honecker:2013mya,Petrossian-Byrne:2025mto}, type IIB~\cite{Cicoli:2012sz,Cicoli:2013ana,Demirtas:2018akl,Hebecker:2018yxs,Mehta:2021pwf,Cicoli:2021gss,Carta:2021uwv,Cicoli:2021tzt,Demirtas:2021gsq,Cicoli:2022fzy,Gendler:2023kjt,Dimastrogiovanni:2023juq,Sheridan:2024vtt,Benabou:2025kgx,Yin:2025amn,Cheng:2025ggf}, the heterotic string~\cite{Choi:2011xt,Choi:2014uaa,Buchbinder:2014qca,Agrawal:2024ejr,Loladze:2025uvf,Leedom:2025mlr,Benabou:2026jtv} as well as M-theory~\cite{Svrcek:2006yi,Im:2019cnl,DeLuca:2021pej} and F-theory~\cite{Halverson:2019cmy,Halverson:2019kna,Fallon:2025lvn,Chen:2026bxp}, they also offer the opportunity to test some of these theories in the lab and in cosmology. 
In addition to explicit string theory constructions, bottom-up approaches to understand the rich phenomenology of the string axiverse have been carried out recently in multi-axion systems \cite{Kitajima:2014xla,Cyncynates:2021xzw,Gavela:2023tzu,deGiorgi:2024elx,Murai:2024nsp,deGiorgi:2025ldc,Baryakhtar:2026oun,Lee:2026umy,FernandezNavarro:2026cyu}, showing how the existence of multiple axions and the interactions among them can have non-trivial implications in cosmology as well as for the experiment. 
In \cite{Baryakhtar:2026oun} it has been argued that under minimal assumptions, in an axiverse with many axions, the QCD axion stands out as the most easily accessible at axion dark matter experiments, a conclusion that aligns with top-down approaches. 
It is worth mentioning that, in addition to testing unification, there has been interest in understanding how the quantization of axion couplings may help us understand the UV limit of the SM~\cite{Reece:2023iqn,Choi:2023pdp,Cordova:2023her,Agrawal:2023sbp,Cordova:2024ypu,Delgado:2024pcv,Choi:2026oqz,Lee:2026umy} with applications ranging from obtaining information about the global structure of the SM to discriminating between pre- and post-inflationary axion cosmology.  

In the context of F-theory GUTs, because hypercharge flux is a topologically non-trivial GUT symmetry breaking mechanism, it is not obvious if axions will satisfy Eq.~\eqref{eq:coupling-to-mass-ratio}. Indeed, it has been shown that turning on a non-trivial hypercharge flux leads to non-universal -- that is, non-GUT symmetric -- holomorphic corrections to the gauge kinetic function~\cite{Blumenhagen:2008aw,Donagi:2008kj,Mayrhofer:2013ara}. Their holomorphic nature implies that similar to the gauge couplings, the axion-gauge boson couplings also receive such corrections. These corrections  lead, in the 4d EFT, to axions that couple in a non-GUT-symmetric way. Similar non-universal corrections appear in heterotic models with non-trivial line bundles \cite{Blumenhagen:2005ga,Blumenhagen:2006ux}. As recently shown in~\cite{Reig:2025dqb}, in such heterotic models the non-universal ALPs still satisfy \eqref{eq:coupling-to-mass-ratio} due to their large mass in the weak coupling regime, $g_s\lesssim O(1)$.

The aim of this work is to  analyse to what extent any axion coupled to photons obeys  Eq.~\eqref{eq:coupling-to-mass-ratio} in F-theory GUTs. Once the holomorphic correction due to the hypercharge flux is taken into account, axions coupled to photons without coupling to gluons appear. We will refer to these axions as non-universal ALPs. We will show that when the threshold corrections to the gauge kinetic functions are small, and the unification of gauge couplings is still approximate as one expects in minimal, phenomenologically viable models, the non-universal ALPs obtain a large mass from stringy non-perturbative effects. More concretely, we will relate the smallness of the threshold correction to the smallness of the D-instanton action that breaks the shift symmetry of the ALP.
Therefore, non-universal F-theory ALPs have a small coupling-to-mass ratio due to their large mass. The more precisely the gauge couplings unify in the UV -- which occurs in the strong $g_s$ limit -- the smaller the ratio $g_{a\gamma}/m_a$ for the non-universal ALPs becomes.\footnote{This conclusion resembles the one obtained for the boundary-localised ALPs in orbifold GUTs~\cite{Agrawal:2025rbr}.}

We will also consider scenarios with large threshold corrections. In these cases, to obtain phenomenologically viable models, new states that do not fill complete GUT representations are required at intermediate energy scales. These states increase the UV value of the gauge couplings. We will see that, in some cases, the UV instanton contribution can be sizeably larger than the IR contribution from QCD, thus spoiling the axion solution to the strong CP problem. By studying explicit examples we will see that, even if one gives up the QCD axion solution to the strong CP problem, a light ALP above the QCD band can only appear in scenarios with small 4-cycles where the $\alpha^\prime$ expansion breaks down. 

\begin{figure}[t]
    \centering  \includegraphics[scale=0.32]{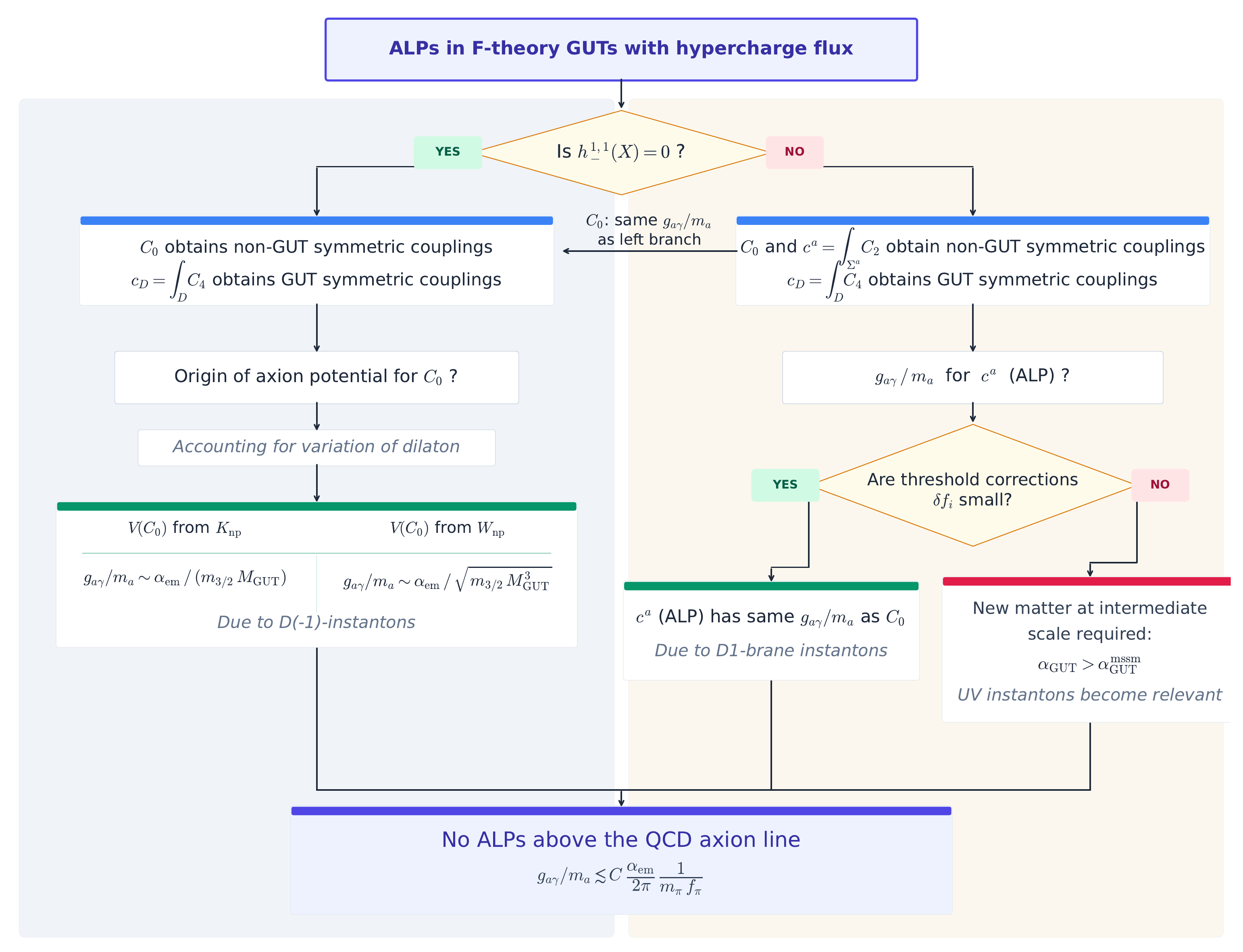}
    \caption{\textbf{Summary of results.} Flowchart summarizing the main arguments leading to the result in Eq.~\eqref{eq:coupling-to-mass-ratio}. It considers the coupling-to-mass ratio for axions from the RR fields (for more details on axions from the complex structure moduli, see Section~\ref{sec:complex_structure_axions}).  The argument assumes a non-vanishing Pfaffian factor, as detailed in Section \ref{sec:Pfaffian_suppression}. Fig.~\ref{fig:ga_vs_ma_plot} contains the F-theory GUT predictions for $g_{a\gamma}$ and $m_a$, together with current and future bounds.}  
	\label{fig:cartoon_diagram}
\end{figure}

\begin{figure}[t!]
    \centering  \includegraphics[scale=0.4]{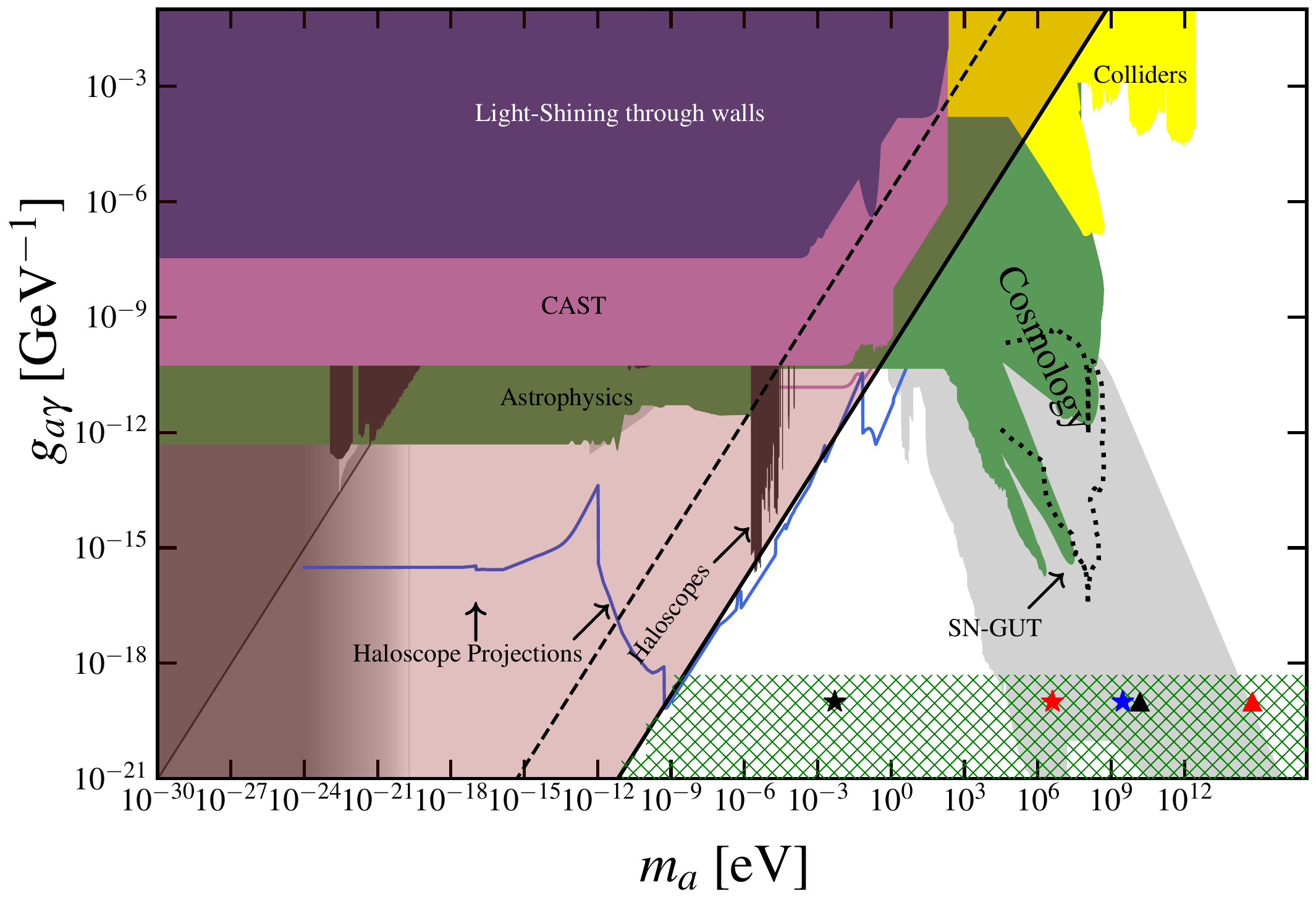}
    \caption{\textbf{Summary of results.} Predictions for $g_{a\gamma}$ and $m_a$ in F-theory GUTs, together with existing experimental constraints and expected sensitivities at future experiments. The shaded red region corresponds to parameter space that is not accessible in F-theory GUTs (assuming a non-vanishing Pfaffian, see Section~\ref{sec:Pfaffian_suppression}) -- finding an axion in this region would rule out this class of theories. This is because in this class of theories every axion satisfies \eqref{eq:coupling-to-mass-ratio}. The stars (triangles) correspond to different benchmark points where the $C_0$ and $c^a$ axion potential comes from non-perturbative corrections to the K\"ahler potential (superpotential). Different colors correspond to different values of the gravitino mass: black corresponds to $m_{3/2}=\mathcal{O}(1)$ eV, \red{red} to $m_{3/2}=\mathcal{O}(1)$ GeV, and \textcolor{blue}{blue} to $m_{3/2}=\mathcal{O}(1)$ TeV. When cosmologically produced via misalignment mechanism, the benchmark points falling in the gray region are excluded from a variety of cosmological constraints. For these benchmarks we take the threshold corrections, and hence the relevant D-instanton actions, to be $\delta\alpha_i^{-1}\sim 10\%$ of $\alpha_{\rm GUT}^{-1}$. As the correction (instanton action) is lowered, each benchmark point displaces to the right. If they are larger, the points move to the left but one has to account for the effect of matter at intermediate scales, which correct the gauge coupling prediction in the UV. As discussed in Section~\ref{sec:large_threshold_and_ga/ma}, also in that case no axion above the QCD line appears, showing the robustness of our result in Eq.~\eqref{eq:coupling-to-mass-ratio}. The hatched region corresponds to the parameter space that can be populated by the $C_0$ and $c^a$ axions for different values of the threshold correction and the gravitino mass, or by other $C_4$ axions that mix with the QCD axion. The black dashed corresponds to the D3-brane-induced ALP potential for $m_{3/2}= 1$ TeV, which is an irreducible contribution that adds to the contributions from D$(-1)$ and D1-instantons. In the (hypothetical) case that the Pfaffian of D$(-1)$ and D1-instantons is vanishingly small, $\mathcal{A}\ll 10^{-66}$, axions between the solid and dashed black line would be allowed, see Section \ref{sec:Pfaffian_suppression}, while the region above the dashed line is excluded. Experimental bounds are adapted from~\cite{AxionLimits}.}
	\label{fig:ga_vs_ma_plot}
\end{figure}

Furthermore, we will also show that including the variation of the dilaton throughout the compact space -- as expected in F-theory constructions -- only strengthens our conclusions. This occurs because in cases with varying $g_s$, the non-universal ALPs generically receive a smaller coupling-to-mass ratio due to the contribution from non-perturbative effects localised in regions of the base where $g_s\sim \mathcal{O}(1)$. Instantons living in this region of the compact space have a small action and generate large shift-symmetry breaking effects. Figure \ref{fig:cartoon_diagram} contains the main logic behind the results, which are summarised in Figure \ref{fig:ga_vs_ma_plot}. 

From a string theoretic point of view, our main observation is that the appearance of instanton generated axion potentials is generic and that the suppression factor can be estimated at least in regions where the effective theory is under control. Indeed, instanton corrections to the hypermultiplet moduli space in 4d ${\cal N} =2$ compactifications are ubiquitous, and we argue that the truncation of the 4d ${\cal N} =2$ corrections found in \cite{Robles-Llana:2006hby,Robles-Llana:2007bbv,Saueressig:2007dr} to the 4d ${\cal N}=1$ effective action results in D$(-1)$ and D1-instanton contributions to at least the K\"ahler potential. The main caveat is the possibility of a small Pfaffian prefactor. This can either be a sign of strong quantum corrections, in the sense stressed more generally in the recent \cite{Kaufmann:2026fli, Kaufmann:2026mha}, hence pointing to a loss of control over the effective theory, or result from vectorlike charged fermionic zero modes becoming light. 
 However, we will see that even in the hypothetical case that the Pfaffian vanishes for all relevant D$(-1)$ and D1-instantons, fluxed D3-brane instantons generate a minimal potential for the ALPs, as expected already on more general grounds such as the Axionic Weak Gravity Conjecture \cite{Arkani-Hamed:2006emk}.

This article is structured as follows. In Section \ref{sec:hypercharge_flux_review} we review the GUT symmetry breaking mechanism by hypercharge flux. In  Section \ref{sec:axion_couplings} we show how, in addition to the QCD axion, this mechanism introduces non-universal ALPs coupled to photons. After that, in Section \ref{sec:g_a/m_a_and_unification} we will estimate the shift symmetry breaking effects for ALPs, relating their strength to the precision of gauge coupling unification. We also consider more general cases where $g_s$ varies along the compact manifold and the effects due to small Pfaffian factors. In Section~\ref{sec:large_threshold_and_ga/ma} we consider scenarios with large threshold corrections. Section \ref{sec:open_string_axion} includes a discussion of cases where there are open string axions and Section \ref{sec:pheno} contains the phenomenological implications for axions in F-theory GUTs. Finally, we conclude in Section \ref{sec:conclusion}.

\section{GUT symmetry breaking by hypercharge flux: a review}
\label{sec:hypercharge_flux_review}

 GUT models in F-theory \cite{Beasley:2008dc,Beasley:2008kw,Donagi:2008ca,Donagi:2008kj} rely on effects which are genuinely non-perturbative in the Type IIB string coupling $g_s$.
As such, the theory is best described in the language of an elliptic Calabi-Yau fourfold $Y_4$ with base $B_3$.\footnote{For background on this geometric language see e.g. the review \cite{Weigand:2018rez} and references therein.}
For example, in a GUT based on gauge group $SU(5)$, the existence of an order-one top Yukawa coupling requires strong coupling effects. These can be understood in the language of exceptional symmetry enhancements over points on the base $B_3$ \cite{Beasley:2008dc,Beasley:2008kw,Donagi:2008ca,Donagi:2008kj},
which are not possible at weak coupling.
Equivalently, these effects can be described as unsuppressed instanton effects \cite{Collinucci:2016hgh,Blumenhagen:2007zk} signalling a departure from small $g_s$ at least in some regions of the compactification space. 
More generally, perturbative and non-perturbative $g_s$ effects are fully resummed in the geometry of $Y_4$.
Its complex structure moduli encode
what in Type IIB language  includes the dilaton controlling $g_s$, the geometric complex structure moduli of the Type IIB Calabi-Yau threefold $X_3$ as well as the open string moduli in the 7-brane sector, all of which are mixed by genuine ${\cal N}=1$ quantum effects. 

The cleanest approach to studying the effective action of F-theory is therefore directly via duality with M-theory compactified on the elliptic fibration $Y_4$ and its uplift to F-theory \cite{Grimm:2010ks}.\footnote{For a recent study of the QCD axion in F-theory along these lines, see \cite{Chen:2026bxp}.} 
For our analysis, however, it will be sufficient, and easier, to describe the model in the language of Type IIB string theory with 7-branes on $B_3$, taking into account $g_s$ effects (including the variation of $g_s$ along $B_3$) explicitly when they become relevant. Oftentimes, we will think of $B_3$ as an orientifold quotient of a Calabi-Yau threefold $X_3$. While this is certainly only fully accurate in a perturbative context, the key properties we are interested in can already be captured by such an analysis, with suitable adjustments. For example, as pointed out already above, the presence of top Yukawa couplings in this language corresponds to singularities on $X_3$ supporting certain D-brane instantons \cite{Collinucci:2016hgh} which leave the strictly perturbative regime. With this understanding, we will treat flux backgrounds directly as gauge fluxes with support on 7-brane divisors and consider axions which arise from the Type IIB Ramond-Ramond gauge potentials $C_0$, $C_2$ and $C_4$ as well as from the Kalb-Ramond 2-form $B_2$. We will also comment on mixings with complex structure moduli as a result of $g_s$ corrections. 

We now briefly review the mechanism of hypercharge flux symmetry breaking in type IIB and F-theory GUTs, which will be important for our analysis. After describing how the GUT gauge group is broken, we will discuss the conditions to avoid a St\"uckelberg mass for $U(1)_Y$, as well as non-universal holomorphic corrections to the gauge kinetic functions. For simplicity we will consider the simplest $SU(5)$ GUT, but the mechanism also applies to larger gauge groups such as $SO(10)$ or exceptional groups. We refer the reader to \cite{Beasley:2008dc,Beasley:2008kw,Donagi:2008ca,Donagi:2008kj,Mayrhofer:2013ara} for more details
 and to \cite{Hayashi:2008ba,Marsano:2009wr,Marsano:2009ym,Weigand:2010wm, Heckman:2010bq,Grimm:2009yu,Dudas:2010zb,Dolan:2011iu,Palti:2012dd,Braun:2014pva,Li:2022aek,Li:2023dya} and references therein for an incomplete list of works highlighting different aspects of GUT constructions in F-theory more generally.

\subsubsection*{Symmetry breaking and the $U(1)_Y$ gauge boson mass}
We consider an $SU(5)$ model given by a stack of 7-branes wrapping a divisor $D$ of the base $B_3$ of the F-theory Calabi-Yau fourfold $Y_4$ \cite{Beasley:2008kw,Donagi:2008ca,Donagi:2008kj,Beasley:2008dc}. 
The main idea of GUT symmetry breaking by a non-trivial hypercharge flux is to embed a line bundle $L_Y$ on $D$ into the $SU(5)$ gauge group by identifying its structure group  with the hypercharge generator, $T_Y$. The hypercharge flux is described by the first Chern class 
\begin{equation}
c_1(L_Y)= \frac{1}{2\pi} F_Y|_D\equiv \mathcal{F}_Y \in H^2(D) \,.
\end{equation}
The total gauge flux on the GUT 7-brane can be written as
\begin{equation} \label{eq:defFflux}
    \mathcal{F} = \mathcal{F}_S\text{diag}(1,1,1,1,1)+ \frac{1}{6} \mathcal{F}_Y\text{diag}(-2,-2,-2,3,3) \,,
\end{equation}
where ${\cal F}_S = \frac{1}{2\pi} F_S|_D$ is the part of the gauge flux commuting with $SU(5)$.\footnote{In the language of Type IIB orientifolds, the generator associated with this gauge background is the diagonal $U(1)$ of $U(5)$ which becomes massive by a geometric St\"uckelberg mechanism \cite{Grimm:2011tb}. It is absent as a selection rule at the non-perturbative level due to instanton effects \cite{Blumenhagen:2007zk} which are automatically encoded in the so-called $E_6$ Yukawa couplings in F-theory \cite{Collinucci:2016hgh}.  For a precise match between the gauge flux in the language of Type IIB orientifolds versus F-theory on elliptic fourfolds we refer to \cite{Krause:2012yh}.  } 
 As a result, the GUT gauge group breaks as
\begin{equation}
    SU(5)\rightarrow SU(3)_C\times SU(2)_w\times U(1)_Y\,.
\end{equation}
The gauge bosons associated to the generators that do not commute with $\mathcal{F}_Y$, i.e. in representation  $R_X\sim ({\bf 3},{\bf 2},\pm 5/6)$, will obtain a mass close to the Kaluza-Klein scale associated with $D$.

A non-trivial hypercharge flux could generate a St\"uckelberg mass for the hypercharge gauge boson. For example, this is what happens in analogous constructions for the heterotic string \cite{Blumenhagen:2005ga,Blumenhagen:2006ux}. On the other hand, Type IIB constructions and more generally F-theory GUTs offer additional structure to avoid this phenomenologically unwanted phenomenon. St\"uckelberg masses for gauge $U(1)$s are derived by dimensionally reducing the Chern-Simons (CS) part of the 7-brane action. This part of the action describes how the branes, and the gauge sectors they host, couple to the different Ramond-Ramond (RR) fields. These RR fields include the fields $C_0$, $C_2$, and $C_4$. These fields can be expanded in terms of the basis of odd and even 2-forms and 4-forms as 
\begin{align}\label{eq:C2_C4_RR_fields}
    C_2 = c^a\omega_a+...\,,\,\, \qquad  C_4 = c_D \tilde \omega^D + c_\alpha \tilde \omega^\alpha + c_2^\alpha\omega_\alpha + ... \, ,
\end{align}
where we single out the four-form  localised to the GUT divisor, $\tilde \omega^D$, and the corresponding axion $c_D$.
Here we view $B_3$ as a $\mathbb Z_2$ quotient of a Calabi-Yau threefold $X_3$, on which
\begin{equation}
    \omega_\alpha\,, \quad \alpha = 1, ... h^{1,1}_+(X_3)\,, \qquad \qquad \omega_a, \quad a = 1, ... h^{1,1}_-(X_3)\,,
\end{equation}
form a basis of the orientifold even and odd co-homology groups $H_\pm^{1,1}(X_3)$, respectively, and 
 $\tilde \omega^\alpha$ is a basis of  even 4-forms $H_+^{2,2}(X_3)$.
  Similarly, the other relevant higher-form field is the Neveu-Schwarz (NS) $B$-field, which can be expanded in terms of orientifold-even and odd two-forms as\footnote{The even part of $B_2$ is projected out by orientifold parity at the level of dynamical fields. Note furthermore that $H^{1,1}(B_3) = H^{1,1}_+(X_3)$, while the modes associated with $H^{1,1}_-(X_3)$ arise, in F-theory, by expanding the M-theory 3-form $C_3$ along elements of $H^3(Y_4)$ with one leg along the fiber \cite{Grimm:2010ks}.}

\begin{equation}\label{eq:B-field}
    B_2 = b^\alpha\omega_\alpha + b^a\omega_a \,.
\end{equation}

Schematically, St\"uckelberg masses for $U(1)$ gauge bosons originate in the dimensionally reduced 4d couplings of $U(1)$ field strengths to 4d two-forms, 
\begin{equation}
    F_i\wedge \text{(2-form)}\,.
\end{equation}
For the hypercharge gauge boson the relevant term is 
\begin{equation} \label{eq-SSt.}
     S_{\rm St}\simeq  {\frac{\mu_7}{8\pi^2}} \int \tr\, T_Y^2 \,F_Y\wedge c_2^\alpha\int_D \mathcal{F}_Y\wedge \omega_\alpha\, ,
\end{equation}
where $\mu_7$ is the 7-brane tension.
 Since $\tr\, T_Y^2 \neq 0$ we note that  a massless $U(1)_Y$ requires that  $\int_D \mathcal{F}_Y\wedge \omega_\alpha = 0$ for every  $\omega_\alpha \in H_+^2(X_3)$. This condition can be realised \cite{Donagi:2008ca, Beasley:2008kw} if the 2-form ${\cal F}_Y$ does not have any components along $\iota^\ast H^2_+(X_3)$, where $\iota: D \to X_3$ is the embedding of the $SU(5)$ GUT divisor $D$ into $X_3$.\footnote{Unless the difference is important, we will not distinguish between the divisor $D$ on the base $B_3$ of $Y_4$ and its uplift to Type IIB on $X_3$. Furthermore, to keep the notation light, we will omit the pullback symbol for $\omega_\alpha$ in expressions like $\int_D \mathcal{F}_Y\wedge \omega_\alpha$.} 
Notably, this leaves room for 
 a hypercharge flux component $\iota^\ast H^2_-(X_3)$ \cite{Mayrhofer:2013ara}, which will become important in the discussion around \eqref{eq:CS_action_in_terms_of_4d_axion}.
 This mechanism relies on the localisation of the GUT group on a divisor $D$ in $B_3$, which explains why it is not available in heterotic models.\footnote{In heterotic compactifications with line bundles, hypercharge can remain massless if it is a linear combination of several individually St\"uckelberg massive abelian subgroups \cite{Blumenhagen:2005ga,Blumenhagen:2006ux}.}

\subsubsection*{Holomorphic corrections to the gauge kinetic function}
Let us now describe the impact of hypercharge flux breaking on the gauge couplings of the 4d EFT. The tree-level $SU(5)$ gauge kinetic function is given by
\begin{equation}\label{eq:GUT_gauge_kin_function}
    f_D = \frac{1}{2g_s l_s^4}\int_{D}J\wedge J+i\int_{D}C_4\,.
\end{equation}
Here $g_s=e^{\phi}$ is the string coupling in terms of the ten-dimensional Type IIB dilaton, $\phi$, and $\text{Vol}(D)=\frac{1}{2}\int_{D}J\wedge J$ is the volume of the divisor around which we wrap the GUT 7-branes. 
Eq.~\eqref{eq:GUT_gauge_kin_function} indicates that the unified gauge coupling is given by the volume of the 4-cycle in string length units, 
\begin{equation}\label{eq:SU(5)_gauge_kin_function}
    \text{Re}[f_D]=\frac{1}{\alpha_{\rm GUT}}=\frac{1}{g_s}\frac{\text{Vol}(D)}{l_s^4}\,.
\end{equation}
The gauge kinetic function $f_D$ also contains the coupling of the RR 4-form field, $C_4$, which in the 4d EFT provides an axion coupled in a GUT symmetric way.

In presence of a non-trivial hypercharge flux, $\mathcal{F}_Y\neq 0$, the gauge kinetic functions of the different SM group factors split up as 
\begin{equation}\label{eq:generic_gauge_kin_function_corrected}
    f_i = \tilde f_D+\delta f_i+\epsilon_i(U)\,,
\end{equation}
where the label $i$ refers to QCD, EW and hypercharge. Here  the $SU(5)$ symmetric contribution, $\tilde f_D$, which includes the GUT-symmetric threshold correction, is modified in a non-universal way by the $\delta f_i$, which depend on the fluxes \eqref{eq:defFflux}, the axio-dilaton and the RR and NS 2-form fields. We have also included the (unknown) contribution from the complex structure moduli, $\epsilon_i(U)$, which can be non-universal and will be discussed in Section \ref{sec:complex_structure_axions}.
 The most general non-universal, holomorphic correction is~\cite{Mayrhofer:2013ara,Jockers:2005zy} 
\begin{equation}\label{eq:gauge_kin_fun_split}
    \delta f_i =i\,\frac{\delta_i}{2} \int_{D}\mathcal{F}_Y\wedge\left [ S(2\tilde{\mathcal{F}_S}+\mathcal{F}_Y) + 2 \left(C_2 - S B_2 \right) \right ]\,,
\end{equation}
where 
\begin{equation} \label{eq:SC2def}
    S=ie^{-\phi}+C_0\,,
\end{equation}
is the complex axio-dilaton. Note furthermore the contribution from
the RR 2-form $C_2$ as well as the $B_2$-field, whose expansion was given in Eq.~\eqref{eq:B-field}. Following \cite{Mayrhofer:2013ara} the flux has been  defined as 
\begin{equation}
    \tilde{\mathcal{F}_S} = \mathcal{F}_S-\frac{2}{5}\mathcal{F}_Y \,.
\end{equation}
The group theoretic coefficients are $\delta_{SU(3)} = 0$, $\delta_{SU(2)} = 1$, $\delta_{U(1)_Y} = 3/5$.
It is also convenient to define the flux-dependent (integer) quantities 
\begin{align}\label{eq:flux_dependent_integers}
    p_{YY} = \int_{D}  \mathcal{F}_Y\wedge \mathcal{F}_Y \,,\hspace{0.7cm}
    p_{YS} = \int_{D}  \mathcal{F}_Y\wedge \tilde{\mathcal{F}}_S \,,\hspace{0.7cm}
    p_{Ya} = \int_{D}  \mathcal{F}_Y\wedge \omega_a  \,.
\end{align}
Again,  $\omega_a$ is a basis of the odd co-homology group, $H_{-}^{1,1}(X_3)$, with $a = 1, ..., h_{-}^{1,1}(X_3)$.

In particular, the real part of Eq.~\eqref{eq:gauge_kin_fun_split} indicates that the gauge couplings receive non-universal corrections,
\begin{align}
	 \label{eq:def_non_univ_threshold}
	 \alpha_i^{-1}&=\alpha_{\rm GUT}^{-1}+\delta\alpha_i^{-1}\,,\\  \label{eq:def_non_univ_threshold-b}
     \text{ with}\,\,\delta\alpha_i^{-1}\equiv \text{Re}\left [\delta f_i\right ] +\text{Re }[\epsilon_i(U)]& 
	 = -\frac{1}{2}\frac{\delta_i}{g_s} \left ( 2p_{YS} + p_{YY} -2 b^a p_{Ya} \right ) +\text{Re }[\epsilon_i(U)] \,,
\end{align}
with $b^a$ being the VEV of the orientifold odd axions from $B_2$. 
The expressions for $\delta f_i$ can be deduced, by holomorphy, from the non-GUT symmetric ALP couplings, which are encoded in ${\rm Im}(f_i)$ and will be derived in Section \ref{sec:axion_couplings}. 
Eq.  \eqref{eq:def_non_univ_threshold-b}  is of particular importance to us, because it will allow us to infer the relation between the non-universal corrections to the gauge couplings and the action of the instantons breaking the shift symmetry of the non-universal axions in Section \ref{sec:g_a/m_a_and_unification}.

Note that F-theory includes the backreaction of $7$-branes into the geometry. These objects (magnetically) source  a non-trivial profile  for the complex dilaton, $S$. As a result, the dilaton field value depends on the position in the internal space, $\phi = \phi(u^i)$, with $u^i$ being internal coordinates. The string coupling $g_s$ can therefore change by $\mathcal{O}(1)$ amounts, becoming very weak in some regions of the internal manifold and diverging in others. This implies that the string coupling appearing in the gauge kinetic functions~\eqref{eq:SU(5)_gauge_kin_function} as well as the threshold corrections~\eqref{eq:gauge_kin_fun_split} corresponds to the averaged value over the divisor or over a given odd 2-cycle. 
Finally, we note that in some specific constructions such as the case of exceptional divisors, $g_s$ is nearly constant and fixed to $\mathcal{O}(1)$ values throughout the divisor~\cite{Dasgupta:1996ij}. For this reason, in F-theory GUTs with exceptional groups like $E_6$, $E_7$ and $E_8$, the string coupling can be taken approximately constant, $g_s\sim \mathcal{O}(1)$. See Appendix \ref{app:varying_gs} for more details. 

\section{Axion couplings in F-theory GUTs}\label{sec:axion_couplings}
The axion couplings to gauge bosons in the 4d EFT are obtained by dimensionally reducing the Chern-Simons (CS) part of the 7-brane action, which describes the coupling of RR fields to the gauge sector on the worldvolume of the 7-branes.
 If we ignore the curvature terms, which play no role in our discussion, the CS action reads 
 \begin{equation}\label{eq:CS}
   S_{\rm CS} = \mu_7  \int \left( C_0 + C_2  + C_4 \right) \wedge e^{ F / 2\pi -B_2} \, , \qquad \mu_7 = \frac{2 \pi}{l_s^8} \,,
\end{equation}
with the understanding that only integrals of 8-form combinations of fields  over the 7-brane worldvolume are non-zero.
This includes the GUT-symmetric coupling of $C_4$ to gauge bosons, 
\begin{equation}\label{eq:CS_C_4}
   S_{\rm CS}\supset  \frac{\mu_7}{8 \pi^2} \int C_4\wedge \tr F^2\,.
\end{equation}
The GUT-symmetric coupling is encoded in the gauge kinetic function, $f_D$, whose universal contribution is given in Eq.~\eqref{eq:GUT_gauge_kin_function}. 
From $f_D$ we also deduce that
\begin{equation}
    c_D = \int_{D} C_4
\end{equation}
is the axion that, before the inclusion of fluxes, couples in a GUT symmetric way to gauge bosons. 

Let us now focus on the non-universal threshold corrections to axion couplings. The non-trivial effects of hypercharge flux on gauge coupling unification come from terms like $\tr F^4$ and $\tr F^3$ in the Dirac-Born-Infeld (DBI) action. These corrections, due to holomorphy, appear together with axions coupled non-universally to gauge bosons. The relevant axionic terms arise from the coupling of other RR fields to the gauge bosons \cite{Blumenhagen:2008aw,Mayrhofer:2013ara},
\begin{equation}\label{eq:non_univ_correction_CS_action}
    S_{\rm CS}\supset 
    \frac{\mu_7}{{4! (2\pi)^4}}\int C_0\wedge \tr F^4 + \frac{\mu_7}{{3! (2\pi)^3}}\int C_2\wedge \tr F^3 - \frac{\mu_7}{{3! (2\pi)^3}}\int C_0\wedge B_2 \wedge \tr F^3\,.
\end{equation}
In models with non-trivial hypercharge flux the terms in Eq.~\eqref{eq:non_univ_correction_CS_action} result in additional ALPs coupled to gauge bosons in a non GUT-symmetric way. Note that the ALPs from $C_2$ and $B_2$ are only relevant in constructions with $h^{1,1}_-(X_3)\neq 0$. In addition to the RR axion $C_0$, these theories have additional ALPs from the $c^a$ and $b^a$ fields defined in \eqref{eq:C2_C4_RR_fields} and \eqref{eq:B-field}. 
The field $C_2$ pairs up with $B_2$ in the combination $C_2 - S B_2$, and when integrated over odd 2-cycles, $\Sigma^a$, give rise to the $\mathcal{N}=1$ complex scalar
\begin{equation}
	\label{eq:Ga_axions}
	G^a = \int_{\Sigma^a}\left(C_2 - S B_2 \right)= c^a-Sb^a\,,
\end{equation}
which transforms appropriately under $SL(2,\mathbb Z)$.

In terms of 4d axion fields and integrals of fluxes over the GUT divisor $D$, as given in \eqref{eq:defFflux},  the above action leads to 
\begin{equation}\label{eq:CS_action_in_terms_of_4d_axion}
    S_{\rm CS} = \frac{\mu_7}{8\pi^2}\int_{M_4}   \tr \left [ F\wedge F \right ] \left( c_D + (c^a -C_0 b^a) \int_D \omega_a\wedge \mathcal{F}  + \frac{C_0}{2} \int_D \tr \left [ \mathcal{F}\wedge \mathcal{F} \right ] \right)\,.
\end{equation}
For a non-vanishing coupling of $c^a-C_0 b^a$, the integral $\int_D \omega_a \wedge \mathcal{F}$, $\omega_a \in H^{1,1}_-(X_3)$, must be non-zero; as stressed after \eqref{eq-SSt.}, this is compatible with the absence of a St\"uckelberg mass for the hypercharge gauge boson. In other words, the odd cycle is part of the GUT divisor wrapped by the 7-branes.

After integrating over the compact dimensions and choosing units such that $l_s = 1$, we obtain the axion couplings to the massless 4d gauge bosons,
\begin{equation}
    \mathcal{L}\subset \frac{{\tilde a}}{4\pi}\left [ k_1F_Y\tilde F_Y+k_2\tr W\tilde W+k_3\tr G\tilde G\right ]+\frac{b}{4\pi}\left [ n_YF_Y\tilde F_Y+n_W\tr W\tilde W\right ]\,,
\end{equation}
with $n_W\equiv \delta_{SU(2)}=1$, $n_Y\equiv \delta_{U(1)_Y}=3/5$ and gauge indices have been omitted for simplicity. The coefficients $k_i$ are the levels of embedding of each SM gauge group factor into the simple GUT group; for $SU(5)$ these correspond to the usual $(k_3,k_2,k_1)=(1,1,5/3)$. In the Lagrangian above we have separated the GUT symmetric axion linear combination, $\tilde a$, which has contributions from $C_4$ as well as from the $C_0$ and $C_2$ fields from the ALP, $b$. The latter is a linear combination of $C_0$ and $c^a$.
In more detail, the two (dimensionless) axion linear combinations are given by 
\begin{align}\label{eq:axion_linear_combs}
    &\tilde a =c_D + \frac{C_0}{2} p_{SS}+(c^a-C_0b^a)p_{Sa}+\text{Im }[\epsilon_0(U)]\,,\\&
    b= C_0(p_{YS}+ \frac{1}{2}p_{YY})+ (c^a-C_0b^a)p_{Ya}+\text{Im }[\epsilon_i(U)]\,,  \label{eq:axion_linear_combs-2}
\end{align}
in terms of the integers defined in Eq.~\eqref{eq:flux_dependent_integers} as well as the integers
\begin{equation}
    p_{SS}=\int_D \tilde{\mathcal{F}}_S\wedge \tilde{\mathcal{F}}_S, \qquad p_{Sa}=\int_D \tilde{\mathcal{F}}_S \wedge \omega_a \,.
\end{equation}
Furthermore, $\text{Im }[\epsilon_0(U)]$ ($\text{Im }[\epsilon_i(U)]$) represent the complex structure axions with universal (non-universal) couplings to gauge bosons.

If there were no additional confining interactions or stringy instantons with a small action, $\tilde a$ would be the QCD axion and $b$ would be an ALP that does not couple to gluons. In the next section, however, we will see that there exist stringy instantons breaking the shift symmetries of $C_0$ and $c^a$ with a small action and we will study the coupling-to-mass ratio for them, $g_{b\gamma}/m_b$. Note that when $C_0$, the $c^a$, and the complex structure axions are integrated out due to a large mass generated by instantons with small action, the entire non-universal ALP linear combination $b$ is effectively integrated out and only the RR 4-form axion, $c_D$, remains at low energies. The goal of this work is to obtain a precise relation between the stringy instanton action and the size of the non-universal threshold corrections, $S_D\sim \text{Re}[\delta f_i]$.

\subsection{Complex Structure moduli: axions and threshold corrections}\label{sec:complex_structure_axions}
We now comment on the potential complex structure dependent contributions to the QCD axion and the ALP candidate, given by the last term in \eqref{eq:axion_linear_combs} and \eqref{eq:axion_linear_combs-2}, respectively.
 Their generic presence can be inferred from the imaginary part of the gauge kinetic function $f_D$.

Indeed, in addition to \eqref{eq:CS_action_in_terms_of_4d_axion}, there are, in general, further corrections to $f_D$ which depend on the complex structure (and 7-brane open string) moduli, see \eqref{eq:generic_gauge_kin_function_corrected}. For mirror symmetric D6-branes in toroidal Type IIA compactifications, the corresponding corrections were explicitly computed in \cite{Blumenhagen:2007ip}; in Type IIA orientifolds, they are K\"ahler moduli dependent, while the tree-level gauge kinetic functions are complex structure moduli dependent.
 In the F-theory analysis of \cite{Donagi:2008kj}, the complex structure dependent thresholds are induced by integrating out certain Kaluza-Klein modes (see also \cite{Conlon:2009qa}). The importance of such corrections for the global structure of the moduli space has been stressed more recently in \cite{Kaufmann:2026fli,Kaufmann:2026mha}.

Close to infinite distance regions in the complex structure moduli space, the imaginary part of a complex structure modulus enjoys a perturbative shift symmetry and can therefore be identified with an axionic field. Away from these asymptotic regimes, the shift symmetry is broken by what on the mirror dual Type IIA side would be the effect of worldsheet instantons. 
  In F-theory, the Type IIB complex structure moduli and the axio-dilaton are treated on the same footing and the regimes where either one of them develops an axionic shift symmetry correspond to different regions of the complex structure moduli space of the F-theory elliptic Calabi-Yau fourfold $Y_4$. In this sense, the separation between both types of moduli and their axions is not natural.

Like for the RR field induced axions, if the complex structure axion is heavy, it can be integrated out and does not modify the definition of the remaining linear combination for the QCD axion or the ALP. The appearance of an additional CS axion is therefore intimately tied to the quest for moduli stabilisation. 
  In the context of the RR axions, this was stressed already in \cite{Conlon:2006tq}, pointing out that a light axion or ALP requires a non-supersymmetric stabilisation of the associated saxion, as for instance is the case for the K\"ahler moduli in the Large Volume Scenario (LVS).
 By contrast, both
 in KKLT  and in LVS, the complex structure moduli are typically stabilised in a supersymmetric way via a Gukov-Vafa-Witten flux superpotential
 \begin{equation}
     W_{\rm flux}=\frac{1}{l_s^2}\int G_3\wedge \Omega\,,
 \end{equation}
where $G_3 = F_3-SH_3$ involves the RR and NS 3-form fluxes, and $\Omega$ is the holomorphic (3,0)-form. In this case, even if the complex structure moduli are stabilised in an asymptotic regime where a sufficiently high quality shift axionic symmetry might exist (prior to taking into account the effect of fluxes), the axion cannot be light without leading also to a light saxion. This is not possible phenomenologically, as such light moduli would mediate long-range forces (more concretely they modify Newton's inverse square law) that are excluded from fifth force search experiments \cite{Tan:2020vpf,Lee:2020zjt}.
 The only option to consider with a light complex structure axion is therefore when the complex structure moduli are stabilised in a non-supersymmetric way such that the saxion is heavy while the axion is sufficiently light. 
  We refer to \cite{Hebecker:2025tui} and references therein for a recent account of such constructions and their challenges.

In cases with non-supersymmetric stabilisation, the threshold corrections to the gauge kinetic function, see Eq.~\eqref{eq:generic_gauge_kin_function_corrected}, impose a strong constraint. Far away from the boundaries of complex structure moduli space, the correction to the gauge kinetic function is expected to be small \cite{Blumenhagen:2007ip}. In this region, the shift symmetry breaking effects from the complex structure dependent part of the K\"ahler potential,
\begin{equation}
    K_{\rm cs} = -\log \int \Omega \wedge \bar\Omega\,,
\end{equation}
give the complex structure axion a mass at least as large as the gravitino mass, so the associated axion is not light.
Conversely, as one moves towards the boundary of moduli space, the complex structure axion may become light. However, in this case the saxion takes large field values and the corrections to the gauge kinetic function, $\epsilon_i(U)$, and hence to the gauge coupling, become large, see~\eqref{eq:def_non_univ_threshold}. In the limit where the saxion goes to infinity, this contribution dominates over the tree-level value set by the size of the GUT 4-cycle.
Since in realistic GUTs with an MSSM-like spectrum the physical gauge coupling is fixed at $\alpha_{\rm GUT} \sim 1/25$ (additional degrees of freedom only increase this value), one cannot move arbitrarily far into the boundary region of moduli space without spoiling the infrared value of the gauge coupling. 
More generally, it has been argued in \cite{Kaufmann:2026fli,Kaufmann:2026mha,Kaufmann:2026tsy} that for finite values of the K\"ahler moduli, the complex structure directions are bounded.

The general conclusion is that light complex structure axions come with large threshold corrections (from the saxions), which is excluded, while small corrections to the gauge coupling from complex structure moduli (corresponding to regions far away from the boundary) make the complex structure axion heavy. Hence, in phenomenologically viable scenarios, we do not expect light axions above the QCD line from the complex structure moduli, which always satisfy~\eqref{eq:coupling-to-mass-ratio}.
Independently of this logic, the complex structure axion shift symmetry is not perfect and broken by the effect dual to Type IIA worldsheet instantons. The same remarks, in fact, apply to the axio-dilaton.  The source of the corrections are D($-1$)-instantons in the Type IIB frame. Like the mirror dual worldsheet instantons of Type IIA, these are summed up in the classical complex structure moduli space of the F-theory fourfold. 

In the rest of this work we will only consider RR axions, as the connection of their mass to gauge coupling unification can be made more quantitatively precise. 
In Section \ref{sec:open_string_axion} we will point out that the main results of this analysis are unchanged by mixing with what in Type IIB language corresponds to open string axions. These are, in fact, examples of complex structure axions of the elliptic fourfold $Y_4$ not inherited from the complex structure deformations of the base $B_3$.
 
\subsection{The QCD axion mass in F-theory GUTs}\label{sec:dec_const_and_QCD_axion}

The decay constant is in general computed from the eigenvalues of the K\"ahler metric for the complex axion superfields, $\frac{\partial^2 K}{\partial T^i\partial T^j}$. Here $K = K (T^i, T^{i\,\dagger})$ is the (dimensionful) K\"ahler potential, a real function of the complex moduli $T^i$. For example, in the case of $C_4$ axions, we have $T^i = \tau^i+i\theta^i$, with $\tau^i$ the modulus fixing the size of the 4-cycle and $\theta^i= \int_{D^i}C_4 $, and similarly for the other types of axions. The kinetic term for the axions is written in terms of the K\"ahler metric as
\begin{equation}
    \mathcal{L}_{\rm kin} = \frac{1}{2}\kappa_{ij}\partial_\mu \theta^i \partial^\mu \theta^j = \frac{1}{4}\frac{\partial^2 K}{\partial \tau^i\partial \tau^j}\partial_\mu \theta^i \partial^\mu \theta^j = \sum_a \frac{(2 \pi  F_a)^2}{2} \partial_\mu \tilde \theta^a \partial^\mu \tilde \theta^a \,.
\end{equation}
 In the last step we went to a diagonal basis of axions 
 with periodicity $\tilde\theta^a \in [0,1]$ to define the axion decay constants $F_a$.

For closed string axions, the decay constant is expected to lie within the range\footnote{The decay constant of the $C_4$ axion can be lowered in the presence of strong warping~\cite{Choi:2003wr,Choi:2025lkg,Choi:2026kxu}. This mechanism may be difficult to implement in F-theory GUTs, where chiral matter is localised along curves on the divisor GUT divisor. Avoiding proton decay would require that matter curves hosting the SM degrees of freedom are located in the region where the energy scales are not warped down.}
\begin{equation}\label{eq:expected_Fa}
    M_s\lesssim F_a\lesssim \frac{\alpha_{\rm GUT}}{\sqrt{8\pi^2}}M_{\rm Pl}\, ,
\end{equation}
where $M_s$ is the string scale and $M_{\rm Pl}$ the 4d Planck scale. The upper bound (which coincides with the model-independent axion of the heterotic string) occurs for example in simple, approximately factorisable\footnote{By this we mean geometries with at least locally a product structure of the form $ X_p\times X_{6-p}$ (or more generally a fibration with fiber $X_p$). It can be shown that on such backgrounds, if $X_p$ contains the gauge theory (e.g. wrapped D(p+3) branes), the axion is around the upper bound in \eqref{eq:expected_Fa}.} 
geometries while the lower bound occurs for more complicated scenarios where $F_a$ lies close to the string scale, see e.g. \cite{Gendler:2023kjt} for the  Kreuzer-Skarke (KS) type IIB GUT axiverse and \cite{Fallon:2025lvn} for a comprehensive, recent study in the context of different F-theory ensembles.\footnote{The upper bound can be slightly enhanced by an $\mathcal{O}(1)$ factor in Swiss-cheese like geometries~\cite{Benabou:2025kgx,Benabou:2026jtv,Sheridan:2024vtt}.} Open string axions, where the axion decay constant can be substantially smaller, will be considered later in Section~\ref{sec:open_string_axion}. Recently, a refined version of the Axionic Weak Gravity Conjecture suggests upper bounds to $F_a$ -- and hence, lower bound to the QCD axion mass -- similar to the one above~\cite{DiUbaldo:2026rly,Maldacena:2026jqd,Etheredge:2026rio}.

Given that a phenomenologically viable F-theory GUT description requires a string scale $M_s\gtrsim M_{\rm GUT}$, this range suggests that the decay constant of the GUT-symmetric axion from $C_4$ is naturally expected to be near the GUT scale, $F_a\approx M_{\rm GUT}$, leading to the QCD axion mass
\begin{equation}
    m_a^{\rm qcd} = \frac{m_\pi f_\pi}{F_a}\approx 0.5\,\text{neV}\,.
\end{equation}
This value roughly coincides with the model-independent axion of the heterotic string~\cite{Svrcek:2006yi}. See also \cite{Benabou:2026jtv} for a more detailed numerical study of the allowed range for the QCD axion mass  within the heterotic string.
Within the KS type IIB axiverse~\cite{Demirtas:2018akl}, it has been shown that retaining field theory unification predicts a mass range for the QCD axion of the form $10^{-11}\text{ eV }\lesssim m_a^{\rm qcd}\lesssim 10^{-8}$ eV, making the sub-$\mu$eV mass region a well-motivated benchmark for experimental searches~\cite{Benabou:2025kgx}. It would be nice to explore whether this range holds in the context of F-theory GUTs, where additional ensembles beyond KS are possible, by performing an analysis along the lines of~\cite{Fallon:2025lvn}. 

Similar to the model-dependent axions in the heterotic string, the decay constants for the ALPs $C_0$ and $c^a- C_0 b^a$ will be comparable to the GUT-symmetric axion, $F_b\sim F_a$. However, as we will see in the next section, their masses are expected to be much heavier than the QCD axion, making them harder to access experimentally. If produced in the early universe, they may have an important impact in cosmology as we will discuss in Section \ref{sec:pheno}.

\section{Shift-symmetry breaking effects and $g_{a\gamma}/m_a$}
\label{sec:g_a/m_a_and_unification}
We now discuss the potential for the axions entering into the definition of the linear ALP combination \eqref{eq:axion_linear_combs-2}. Axion potentials are generated by non-perturbative effects which break the axionic shift symmetries. In the present context, the first relevant source are D$(-1)$-instantons, which are fully localised in ten dimensions and generate a potential for $C_0$ of the form 
\begin{align}
	\label{eq:V_D(-1)}
	V_{D(-1)} \sim - \Lambda^4 e^{-S_{D(-1)}} \cos \left(2 \pi C_0 \right) \, .
\end{align}
In compactifications with orientifold-odd 2-cycles there are also contributions from D1-branes wrapping these cycles. They generate a potential for the axions $\text{Re}[G^a] = c^a-b^aC_0$ (see eq.~\eqref{eq:Ga_axions}), of the form
\begin{equation}
	\label{eq:V_D1}
	V_{D1} \sim  -\Lambda^4 e^{-S_{D1}} \cos \left(2 \pi  (c^a-b^aC_0) \right) \, .
\end{equation}
  Finally, among the contributions from D3-brane instantons which are of special interest for us are those wrapped on the GUT cycle $D$, where we have to sum over all instanton fluxes 
 as in \cite{Grimm:2011dj}. Depending on the instanton flux, these can take, in particular, the role of the $SU(3)_c$ or $SU(2)_w$ gauge instantons. As we will see, modulo potential vanishings of the Pfaffian prefactors, these instantons are generically more strongly suppressed than the D$(-1)$ and D1-instantons. They are therefore relevant for the ALPs only as a lower bound on the potential in situations where the D$(-1)$ and  D1-instantons are strongly suppressed, as will be discussed in Section \ref{sec:Pfaffian_suppression}.

The instanton actions $S_{D(-1)}, \, S_{D1}$ can be directly related to the threshold corrections,  $\delta\alpha^{-1}_i$, which according to Eq.~\eqref{eq:def_non_univ_threshold} split the gauge kinetic functions of SM couplings in the UV as\footnote{Note that in addition to the fluxes, the splitting depends on the VEV of the orientifold-odd axions, $b^a$. This axion can be stabilised, for example, via a D-term potential \cite{Jockers:2005zy,Grimm:2011dj}.}
\begin{equation}
	 \label{eq:non_univ_threshold}
     \delta\alpha_i^{-1}\equiv \text{Re}\left [\delta f_i\right ] 
	 = -\frac{1}{2}\frac{1}{g_s}\delta_i \left ( 2p_{YS} + p_{YY} -2 b^a p_{Ya} \right ) \,.
\end{equation}
The general picture is therefore that models where the threshold corrections are small typically lead to heavy ALPs, as the instantons are unsuppressed and the contributions to the potential become large. We discuss the possibilities in detail in Sections~\ref{sec:D-1_instantons} and~\ref{sec:D1_instantons}, also taking into account a variation of $g_s$ over the internal manifold. As we will see, regions of large $g_s$ tend to dominate the instanton potential, again leading to substantial contributions to the potential and heavy axions. This is the case for the non-perturbative corrections to the K\"ahler potential and, under some circumstances, for the non-perturbative contributions to the superpotential. 

In a 4d ${\cal N}=1$ compactification, the type of terms to which a class of instantons contributes depends on its zero modes, in particular its fermionic zero modes.
A contribution to the superpotential requires a BPS instanton with exactly two universal fermionic zero modes, while all others must either be lifted through background effects such as fluxes or it must be possible to absorb them in the effective action without inducing a higher derivative effective interaction \cite{Beasley:2005iu}.
For a superpotential term, the resulting contribution \eqref{eq:V_D(-1)} and \eqref{eq:V_D1} to the effective action comes with a suppression from the 
breaking of $\mathcal{N}=2$ SUSY, leading to 
\begin{equation} \label{eq:LambdaWfactor}
    \Lambda_W^4 = \mathcal{A}\,m_{3/2} M_{\rm GUT}^3 \,,
\end{equation}
with $\mathcal{A}$ the one-loop Pfaffian. The Pfaffian depends on different moduli fields and its exact form is generally unknown.\footnote{See \cite{Alexandrov:2022mmy} and references therein for a calculation of this one-loop determinant, $\mathcal{A}$.} It includes both the effects from integrating out heavy string modes and also running effects from light instanton modes. In particular, a vanishing of  ${\cal A}$ can occur when fermionic zero modes become massless and are unsaturated in the Grassmann integral over all instanton modes. We will come back to this question in Section \ref{sec:Pfaffian_suppression}.

 For a K\"ahler potential contribution at the 2-derivative level, the instanton must have four universal zero modes that are not lifted by interactions. Otherwise an instanton contributes higher derivative terms \cite{Beasley:2005iu}.
  The resulting potential is suppressed by an extra factor of $m_{3/2}/M_{\rm GUT}$, leading to 
  \begin{equation} \label{eq:Kaehlersuppression}
  \Lambda_K^4 = {\cal A}_K m^2_{3/2} M_{\rm GUT}^2 \,,
  \end{equation}
  again with a loop-generated moduli dependent prefactor ${\cal A}_K$. In general it will take a different functional form than the Pfaffian in the superpotential, but can also vanish (or become very small) along special loci in moduli space where fermionic instanton modes becomes (approximately) massless.

In Section \ref{sec_Kaehlernonp}, we will argue for the appearance of a K\"ahler potential contribution from both D$(-1)$ and D1-instantons, and comment on the corresponding superpotential in Section \ref{sec:superpotentialgen}. In the following sections we then estimate the suppression of the resulting non-perturbative axion potentials, assuming first that the Pfaffian ${\cal A}$ and its K\"ahler potential analogue ${\cal A}_K$ are of ${\cal O}(1)$.
This allows us to derive bounds on the axion-photon coupling to mass ratio, $g_{a \gamma} / m_a$. Representative examples where the instantons contribute to the superpotential are shown in Fig.~\ref{fig:ga/ma_vs_action_from_Wnp_new}, and in Fig.~\ref{fig:ga/ma_vs_action_from_Knp_new} for K\"ahler potential contributions.
 In Section \ref{sec:Pfaffian_suppression} we address the caveat of very small values for ${\cal A}$ or
${\cal A}_K$ by estimating the effect of otherwise subleading D3-brane instantons.

\subsection{Non-perturbative corrections to the K\"ahler potential } \label{sec_Kaehlernonp}

A mass for axions can be generated already via corrections to the tree-level K\"ahler potential,
\begin{equation}\label{eq:corrected_Kahler_pot}
	\mathcal{K}=\mathcal{K}_{\rm tree}+\mathcal{K}_{\rm pert}+\mathcal{K}_{\rm np}\,.
\end{equation}
Here, $\mathcal{K}_{\rm pert}$ encodes perturbative corrections in $g_s$ and $\alpha^\prime$, which do not introduce any axion dependence, but can stabilize some of the moduli.\footnote{If sizeable, these corrections may change the tree-level value of the decay constant.} By contrast, the non-perturbative corrections, $\mathcal{K}_{\rm np}$, introduce a dependence on the axion field.
In the presence of a non-vanishing superpotential, e.g. from fluxes, these corrections induce a mass term for the axions,  see e.g. \cite{Conlon:2006tq} for a discussion in the Type IIB orientifold context. 

Apart from broader genericity arguments, the appearance of such corrections is suggested already by viewing the 4d ${\cal N}=1$ effective action as an orientifold quotient of an ${\cal N}=2$ theory.
In this parent theory, the axio-dilaton and the K\"ahler moduli are part of the hypermultiplet moduli space, which is known to be substantially corrected by D$(-1)$ and D1-brane instantons. 
As reviewed in more detail in Appendix \ref{app:NP_corrections_to_K}, the ${\cal N}=2$  K\"ahler potential receives perturbative loop and non-perturbative corrections. The non-perturbative corrections include contributions from D$(-1)$ and D1-branes which are required to complete both the 1-loop perturbative correction and the correction due to worldsheet instantons into an $SL(2, \mathbb Z) $ invariant expression~\cite{Robles-Llana:2006hby,Robles-Llana:2007bbv}. The $SL(2, \mathbb Z) $ completion of the worldsheet instanton correction corresponds to $(p,q)$-strings wrapping two-cycles, whose multiplicity is counted by the Gopakumar-Vafa invariants.

To pass to the 4d ${\cal N}=1$ theory, the hypermultiplet moduli are subject to the orientifold projection.  The surviving fields organise in 4d ${\cal N}=1$ chiral multiplets which couple to  instantons in the ${\cal N}=1$ theory.
 The D$(-1)$-instantons are mapped to D$(-1)$-instantons by the orientifold involution and therefore survive the orientifold projection. 
  By contrast, a D1-brane instanton on a curve $C$ is mapped to an anti-D1-instanton along the curve $C'$, which is the image of $C$ with respect to the geometric part of the orientifold involution. 
  Equivalently, the latter can be expressed as a D1-brane instanton along a curve in the class $-C'$ (with the minus sign denoting orientation reversal). In total, the invariant combination is therefore a D1-instanton on the curve $\Sigma = C-C'$, which defines an orientifold odd homology class. Sometimes we will refer to this instanton as a $D1-\overline{D1}$ instanton pair.

  While in 4d ${\cal N}=2$ theories, the moduli space factorises into a direct product of vector- and hypermultiplet moduli spaces,
  this factorisation ceases to hold after the orientifold projection for the ${\cal N}=1$ descendants of the moduli.  
   Nonetheless, the 4d ${\cal N}=1$ effective action contains the 4d ${\cal N}=2$ non-perturbative corrections, suitably projected to the surviving moduli, as a subsector (plus a plethora of additional, genuinely ${\cal N}=1$ corrections that are not inherited from the parent theory). 
    This is true unless the backreaction of genuine ${\cal N}=1$ effects 
  removes the instanton contribution to the K\"ahler potential, essentially by changing its zero mode structure.

  The first such type of effects can occur if the presence of spacetime-filling D-branes changes the structure of the fermionic universal zero modes of the instanton, i.e. 
  the Goldstinos associated with the breaking of the local supersymmetry of the background by the instanton. For an instanton at generic position with respect to the D-branes or the O-planes, the number of Goldstinos is the same as for an instanton in the parent in 4d ${\cal N}=2$ theory. However, the number of 
  universal zero modes is reduced when 
  an instanton lies on top of a D-brane \cite{Petersson:2007sc} or when part of the universal zero modes are projected out by the orientifold action. More details about both effects can be found in the review \cite{Blumenhagen:2009qh} and references therein. Either way, since the number of universal zero modes is only reduced, an instanton contribution to the K\"ahler potential is replaced by a contribution to an F-term, and hence a superpotential. 

 A second type of effect is that the presence of spacetime-filling D-branes can induce additional, charged  fermionic instanton zero modes from open strings stretched between the instanton and the background D-brane.
  There are two qualitatively different scenarios:
  If there is a chiral excess of such charged zero modes, the instanton gives rise only to a charged operator to the effective action \cite{Blumenhagen:2006xt,Ibanez:2006da} and therefore cannot contribute to the axion potential. 
  On the other hand, if the charged zero modes come as non-chiral pairs, they can in principle be saturated in the Grassmann integral. In particular, if the zero modes obtain a mass, either for suitable values of the moduli or after supersymmetry breaking, this suppresses the instanton term by powers of the mass. 
  In the supersymmetric theory, this effect is encoded in the moduli dependent loop generated prefactor ${\cal A}_K$, which vanishes on the loci in moduli space where the charged zero modes are massless. 
  To structure the discussion, we will assume for the time being that ${\cal A}_K = {\cal O}(1)$, corresponding, in particular, to the absence of light charged vectorlike pairs of zero modes, and return to situations where ${\cal A}_K \ll 1$ in Section \ref{sec:Pfaffian_suppression}.

    In the remainder of this section we  argue that the first scenario, of chiral charged fermionic zero modes, does not prevent the generation of a K\"ahler potential from D$(-1)$ and D1-instantons coupling to light axions. 
  Indeed, 
   the presence of D7/D3-branes or O7/O3-planes in the orientifold does not induce any chiral zero-modes on D$(-1)$-instantons or on the relevant  D1-instantons which would nullify their contribution to the effective action inherited from the 4d ${\cal N}=2$ parent theory. For D$(-1)$-instantons this is clear because the open strings from D$(-1)$ to D7-branes (or D3-branes) at best give rise to vector-like pairs of zero modes. 
   By contrast, a single D1-instanton wrapping a general curve of the form $\Sigma = C-C'$ {\it can} carry chiral  fermionic zero modes
   arising from open strings at the intersection with D7-branes wrapping a divisor $D$. 
    Such modes are charged under the diagonal $U(1)$ gauge group and, if present, would  prevent the generation of an axion mass term. However, the presence of these chiral charged zero modes is equivalent to a gauge transformation of the instanton suppression factor because the axionic shift symmetry is gauged by a St\"uckelberg mechanism.
Therefore, charged chiral zero modes only affect D1-instantons that couple to combinations of axions which have St\"uckelberg masses. This effect is therefore irrelevant for our analysis of candidates for light ALPs. Including open string axions does not alter this conclusion, as elaborated in Section \ref{sec:open_string_axion}.

To see this in more detail, recall, e.g. from \cite{Grimm:2011dj}, that the basis of axion fields $G^a$ introduced in \eqref{eq:Ga_axions} carries linear charge $C^a_A$ with respect to the diagonal $U(1)_A$
 gauge symmetry associated with a stack of $7$-branes along a divisor $D_A$ and its orientifold image. Here $C^a_A$ is the coefficient in the expansion of the divisor class $D_A = C^\alpha_A \omega_\alpha + C^a_A \omega_a$ into a basis of $H^{1,1}_\pm(X_3)$. If we expand the curve $C- C' = q_a \Sigma^a$ into a basis of $H_2 ^-(X_3)$, the instanton suppression reads $ e^{2 \pi i q_a G^a}$. This suppression factor shifts under a $U(1)_A$ transformation with gauge transformation parameter $\Lambda^A$
 \begin{equation}
e^{-2 \pi i q_a G^a} \longrightarrow e^{-2 \pi i q_a (G^a + C^a_A \Lambda^A)} \,.
 \end{equation}
This makes it clear that the suppression factor of an instanton coupling to the linear combination $q_a G^a$ is gauge invariant by itself (and hence the instanton carries no chiral charged zero modes) if and only if $q_a C^a_A =0$ for all $A$. This, in turn, is equivalent to $q_a G^a$ not receiving a St\"uckelberg mass (because the massless axions are those in the kernel of the charge mass matrix $C^a_A$). Therefore, all linear combinations which are candidates for a light ALP couple to a gauge invariant instanton without charged chiral zero modes. These instantons are of the type that can be inherited from the underlying ${\cal N}=2$ theory.

To summarise, the structure of non-perturbative corrections to general 4d ${\cal N}=2$ Calabi-Yau compactifications implies an inherited axion potential from non-perturbative corrections to the 4d ${\cal N}=1$ K\"ahler potential, at least in
 scenarios where the D$(-1)$ and D1-brane contributions to the superpotential are strongly suppressed or vanish.\footnote{As discussed above, if two of the four universal zero modes are lifted, a superpotential can replace the K\"ahler potential naively expected to be inherited from the 4d ${\cal N}=2$ theory.} This is expected for both the $C_0$ and $c^a$ ALPs.
 For the latter, since the K\"ahler potential  is a real function, these corrections are schematically of the type
\begin{equation}\label{eq:non_pert_Kahler_Ga}
     \mathcal{K}_{\rm np}\sim e^{-2\pi i q_a^j G^a} + e^{2\pi i q_a^j \bar{G}^a}\,.
\end{equation}
This form of the K\"ahler potential is in agreement with the 4d ${\cal N}=2$ corrections (see Appendix \ref{app:NP_corrections_to_K} for a derivation), which then gives rise to a scalar potential of the form \eqref{eq:V_D1} (in the presence of a non-zero superpotential $W_0$). Similar reasoning leads to D$(-1)$-induced corrections of the form \eqref{eq:V_D(-1)}. In fact, complementary to these arguments based on $SL(2,\mathbb Z)$ invariance in the parent $\mathcal{N}=2$ theory, the generic appearance of D$(-1)$ corrections to the effective action is strongly suggested from the point of view of F-theory, where the classical geometry is known to sum up perturbative and non-perturbative $g_s$ corrections.

In general, all these non-perturbative corrections to the moduli space geometry give rise to an axion potential with an overall scale
\begin{equation}\label{eq:Lambda4_Kahler_potential}
    \Lambda^4  = {\cal A}_K m_{3/2}^2 M_{\rm GUT}^2\,.
\end{equation}
The caveat of a very small or vanishing loop-prefactor ${\cal A}_K$ will be addressed in Section \ref{sec:Pfaffian_suppression}.
 Assuming until then ${\cal A}_K = {\cal O}(1)$, the results in the case that these corrections constitute the leading contribution to the axion potential are shown in Fig.~\ref{fig:ga/ma_vs_action_from_Knp_new}. For the $C_0$ axion, the fact that these corrections do not vanish will be particularly relevant when we consider the position dependence of the dilaton along the compact space.

\subsection{Non-perturbative contributions to the superpotential} \label{sec:superpotentialgen}
 
Interestingly, the generation of a K\"ahler potential in the 4d ${\cal N}=2$ setting reviewed in Appendix~\ref{app:NP_corrections_to_K} does not rely on a specific structure of the (uncharged non-universal) instanton zero modes. This suggests that in the 4d ${\cal N}=1$ context, even a contribution to the superpotential may be more generic than one might at first sight expect; the reason is that interactions in the instanton effective can absorb additional zero modes which would naively prevent a superpotential generation. For a recent discussion along these lines, see \cite{Palti:2020qlc}.
If a superpotential is generated, the instanton suppression is of the form
\begin{equation}\label{eq:Lambda4_superpotential}
    \Lambda_W^4  = {\cal A} \,  m_{3/2} M_{\rm GUT}^3\,,
\end{equation}
with the moduli dependent one-loop Pfaffian
${\cal A}$ introduced already after \eqref{eq:LambdaWfactor}.

The gravitino mass $m_{3/2}$ can take different values depending on the susy breaking mechanism. While gravity mediated supersymmetry breaking suggests that the gravitino mass is comparable to the superpartner masses, 
\begin{equation}
m_{3/2}\sim m_{\rm susy}\gtrsim \mathcal{O}(1) \,  {\rm TeV}\,, \qquad \qquad {\rm (gravity \, \, \, mediation)}
\end{equation}
these two can be decoupled in gauge mediated supersymmetry breaking, where $m_{3/2}$ could be very light~\cite{Giudice:1998bp}. We acknowledge that realizing gauge mediated SUSY breaking within F-theory GUTs, where $M_s\gtrsim M_{\rm GUT}$, can be extremely difficult to achieve in practice, taking into account the need for moduli stabilisation.  This is because the mass of the messenger fields -- open strings stretched between the hidden sector where supersymmetry breaking takes place and the GUT brane -- tends to be comparable to the string or 4d Planck scale. For example, to obtain a very light gravitino mass,  $m_{3/2}=\frac{F}{M_{\rm Pl}}\sim 1$ eV, while maintaining superpartner masses in compatibility with LHC bounds, $m_{\rm susy}\sim \frac{F}{M_{\rm mess}}\gtrsim$ TeV, requires a light messenger field with $M_{\rm mess}\lesssim 10^6$ GeV. This is at least 10 orders of magnitude below the string scale in the models we consider, $M_s\gtrsim M_{\rm GUT}$. When $M_{\rm mess}\sim M_s$, we expect the gravitino mass to be comparable to the other superpartner masses. See \cite{Marsano:2008jq} for an example of light gravitinos within F-theory GUTs, where $M_{\rm mess}\sim 10^{14}$ GeV and $m_{3/2}\sim \mathcal{O}(1)$ GeV, and \cite{Heckman:2008qt} for a gauge mediation scenario with $M_{\rm mess} \sim 10^{12}$ GeV.

\subsection{D$(-1)$-instanton action} 
\label{sec:D-1_instantons}

Let us study now the non-perturbative potential generated by the D$(-1)$-instantons, which generate an axion potential of the form~\eqref{eq:V_D(-1)}, 
\begin{align}
	V_{D(-1)} \sim - \Lambda^4 e^{-S_{D(-1)}} \cos \left(2 \pi C_0 \right) \, .
\end{align}
Here $\Lambda^4$ is given by Eq.~\eqref{eq:Lambda4_superpotential} or \eqref{eq:Lambda4_Kahler_potential} when the D$(-1)$-instantons contribute to the superpotential and the K\"ahler potential, respectively. 

The size of the D$(-1)$-instanton action can be related to the threshold corrections in eq.~\eqref{eq:non_univ_threshold}, as\footnote{\label{footnote:CSaxion_Pfaffian} Here we are ignoring potential complex structure dependent quantum corrections to $S_{D(-1)}$ of the form discussed in \cite{Kaufmann:2026fli}. These are the analogue of the corrections $\epsilon_i(U)$ appearing in the gauge kinetic functions of spacetime filling D-branes, see \eqref{eq:generic_gauge_kin_function_corrected}.  
 For the instanton, they enter the definition of the Pfaffian-like loop prefactor ${\cal A}$ or ${\cal A}_K$. Ignoring these corrections is therefore consistent with our current assumption that ${\cal A}_{(K)} = {\cal O}(1)$, which we will relax at the end of the section. In fact, a substantial loop-correction would likely imply quantum corrections of similar magnitude also for the gauge kinetic functions and be in tension with control of the effective field theory. See also the discussion in Section \ref{sec:complex_structure_axions}.}

\begin{equation}
	\label{eq:threshold_and_instanton_action}
	S_{D(-1)} = \frac{2\pi}{ g_s}=\frac{2\pi\delta \alpha_i^{-1} }{\delta_i}\frac{-2}{2p_{YS}+p_{YY}-2b^ap_{Ya}}\,.
	\end{equation}
Avoiding (vector-like) massless exotics in the $(\mathbf{3},\mathbf{2},5/6)$ representation requires $p_{YY}= - 2$~\cite{Beasley:2008kw,Donagi:2008kj}, 
which implies that for small threshold corrections, $\delta\alpha^{-1}_i\ll \alpha_{\rm GUT}^{-1}\sim 24$, the D$(-1)$-instanton action is also expected to be small,
\begin{equation}
	\label{eq:D-1_inst_small_thresholds}
	\frac{ \delta \alpha_i^{-1} }{\delta_i}\ll \alpha_{\rm GUT}^{-1} \rightarrow S_{D(-1)}\ll \frac{2\pi}{\alpha_{\rm GUT}}\approx 150\,,
\end{equation}
leading to large masses for the RR axion, $C_0$.

In the fine tuned case in which $2p_{YS}+p_{YY}-2b^ap_{Ya}\ll 1$, the above equation indicates that $S_{D(-1)}=\frac{2\pi}{g_s}\gg \delta\alpha_i^{-1}$. As we will see below, in this case the variation of the dilaton throughout the base implies that there will exist regions where $S_{D(-1)}\sim \mathcal{O}(1)$ that will dominate the $C_0$ axion potential.
 
We note that in the geometric regime, when the $\alpha^\prime$ expansion is under control, the action of D$(-1)$-branes is parametrically smaller than the action of D3-branes, $S_{D3} = 2\pi/\alpha_{\rm GUT}$, by a factor proportional to the 4-cycle volume, $\text{Vol}(D)/l_s^4$. We expect this to happen for a string coupling satisfying $g_s \gg \alpha_{\rm GUT}$, see Eq.~\eqref{eq:SU(5)_gauge_kin_function}. This implies that the GUT-symmetric axion $c_D=\int_{D}C_4$ does not receive large shift-symmetry breaking from stringy instantons and in the absence of couplings to other confining interactions it is light enough to become the QCD axion and solve the strong CP problem. 

\subsubsection{Effect of a varying $g_s$}\label{sec:varying_gs_D(-1)}

In general F-theory models, the string coupling $g_s$ need not be a constant throughout the compactification manifold.\footnote{We review this phenomenon and discuss specific examples where the variation of $g_s$ can be calculated in Appendix~\ref{app:varying_gs}.} Since the D-instanton action scales as $1/g_s$, the size of the instanton potential depends on whether the contributing instantons are localised in regions where $g_s$ is large or small. 
  All instanton contributions must be summed up so the dominant contributions will arise from regions where the string coupling is largest, $g_s = g_s^{\rm max}$. This is particularly relevant in phenomenologically viable F-theory GUTs, where the string coupling must be $g_s^{\rm max}\sim O(1)$~\cite{Beasley:2008kw,Donagi:2008kj} in some regions along the GUT divisor, for example along certain matter curves, or in other regions of the base.

Away from the GUT divisor, the variation of the dilaton is smooth, so generically there exists a finite region of the compactification where $g_s$ remains of order one. D$(-1)$-brane contributions localised in such regions therefore generate instanton effects that are effectively unsuppressed, that is with instanton action $S_{D(-1)}\sim \mathcal{O}(1)$. In these cases, the resulting RR-axion mass is 
\begin{equation}\label{eq:C0_mass_varying_gs}
	m_{C_0}^2 \sim  \Lambda^4/M_s^2 \, .
\end{equation}
Here we are taking, for simplicity, the $C_0$ axion decay constant to be at the string scale, and the instanton scale $\Lambda^4$ is given by  Eq.~\eqref{eq:Lambda4_Kahler_potential} or Eq.~\eqref{eq:Lambda4_superpotential} depending on the origin of the axion potential. Thus, in compactifications where the string coupling takes values $g_s\sim \mathcal{O}(1)$ over a finite region -- as is typical in viable F-theory GUT models -- D$(-1)$-instanton corrections to the potential will generate a mass for the $C_0$ ALP comparable to or larger than the gravitino mass, $m_{C_0}\gtrsim m_{3/2}$. This mass will be enhanced in cases where the superpotential contribution is also unsuppressed, see Eq.~\eqref{eq:Lambda4_superpotential}, but we acknowledge that this relies on stabilizing D$(-1)$-branes near regions with $g_s\sim 1$.

\subsection{D1-instanton action}
\label{sec:D1_instantons}
We now discuss the additional axions from expansion of $C_2-C_0 B_2$ along orientifold odd 2-cycles and their D1-instanton induced potential~\eqref{eq:V_D1},
\begin{equation}\label{eq:V_D1_potential}
	V_{D1} \sim  -\Lambda^4 e^{-S_{D1}} \cos \left(2 \pi  (c^a-b^aC_0) \right) \, .
\end{equation} 
In the orientifold limit, the instanton action of a D1-brane wrapping the $a$-th odd cycle is given by
\begin{equation}\label{eq:odd_inst_action}
    	S_{D1}^{a} = \frac{2\pi}{g_s} b^a \,.
\end{equation}
 This is because for an orientifold odd two-cycle, the integral of the K\"ahler form vanishes, $\int_{\Sigma^a}J=0$, and for the instanton with vanishing instanton gauge flux along the two-cycle, the volume form  only involves the odd axions from the $B$-field, \begin{equation} \label{eq:volSigma}
 \text{Vol}(\Sigma^a) =\int_{\Sigma^a} \sqrt{\text{det}(P[g+B]+2\pi\alpha^\prime F)}=  b^a \,.
 \end{equation}
As the VEV of the orientifold-odd axion can be taken to lie in the interval $b^a \in [0,1]$, we find a bound on the instanton action
\begin{equation}\label{eq:relation_inst_actions}
	S_{D1}^a \precsim S_{D(-1)}\,.
\end{equation}
Similar to the case of the RR axion $C_0$, scenarios for which D1-branes generate a superpotential for $c^a$ are also described by Fig.~\ref{fig:ga/ma_vs_action_from_Wnp_new}. On the other hand, when the axion potential comes from non-perturbative corrections to the K\"ahler potential  the coupling-to-mass ratio for $c^a$ axions is given in Fig.~\ref{fig:ga/ma_vs_action_from_Knp_new}.\footnote{Strictly speaking, the potential $V_{D1}$ gives a mass to the combination $c^a-b^aC_0$. However, as we saw above, the axion $C_0$ gets a large mass from effects localised in regions where $g_s$ is large. After integrating it out, $V_{D1}$ gives a mass to $c^a$.}

In either case we find that in the limit of approximately constant string coupling and small holomorphic threshold corrections all the non-universal ALPs, including the RR axion $C_0$ and the 2-form axions $c^a$, satisfy the relation~\eqref{eq:coupling-to-mass-ratio} due to a heavy axion mass.

\subsubsection{Effect of varying $g_s$}\label{sec:varying_gs_D1}

We now estimate $S_{D1}$ in scenarios with varying $g_s$ by taking the orientifold limit. When the manifold has odd 2-cycles one has to consider situations where the orientifold plane intersects with the GUT divisor. We note that for the $c^a$ axion to couple to the  gauge bosons, the odd cycle has to be part of the divisor (see Eq.~\eqref{eq:CS_action_in_terms_of_4d_axion}), so we restrict ourselves to estimating the action of a D1-brane instanton wrapping an odd cycle on the GUT divisor. Axions from $C_2$ integrated over odd cycles which are not on the GUT divisor can still couple to photons via mixing with $C_0$, $c_D$ and $c^a$, however the mixing suppression implies that they satisfy Eq.~\eqref{eq:coupling-to-mass-ratio}.

The gauge coupling, or equivalently the D3-brane instanton action, depends on the averaged value of $g_s$ over the entire GUT divisor, $g_s^{D}$, defined via
\begin{equation}\label{eq:def_gut_coupling_varying_dilaton}
    \frac{1}{\alpha_{\rm GUT}} = \int_D \text{Im}(S)\, \text{dVol} =:\frac{1}{g_s^{D}} \frac{\text{Vol}(D)}{l_s^4}\,.
\end{equation}
Similarly, the action of a D1-brane wrapping an odd 2-cycle depends on the averaged value, called $g_s^a$, of the string coupling along the odd 2-cycle, defined as
\begin{align}\label{eq:g_s_odd_cycle}
   g_s^{a} := \text{Vol}(\Sigma^a) \left[ \int_{\Sigma^a} \text{Im}(S)\, \text{dVol} \right]^{-1} \,.
\end{align}
Here ${\rm Vol}(\Sigma^a)$ is given by \eqref{eq:volSigma}.
 Note that the averaged value $g_s^{a}$ can in principle be different for each of the odd cycles if $h^{1,1}_-(X_3)>1$  (see Appendix \ref{app:odd_cycles} for a discussion of this case).

First, suppose the value of the average is $g_s^{a}\sim \mathcal O(1)$. In this case, as for the D$(-1)$-instanton action, the axion potential from D1-branes around the $a$-th odd cycle will be unsuppressed, $S_{D1}^{a}\sim \mathcal O(1)$, with the axion mass given by $m_{c^a}^2\sim \Lambda^4/F_a^2$, and the scale $\Lambda^4=m_{3/2}M_{\rm GUT}^3$ or $\Lambda^4=m_{3/2}^2M_{\rm GUT}^2$ depending on the origin of the axion potential. 

If the string coupling along the odd 2-cycles is small compared to $g_s^{D}$ (see Eq.~\eqref{eq:def_gut_coupling_varying_dilaton}), $g_s^{a}\ll g_s^{D}$, $c^a$ could become light. However, there is a limit in how small  $g_s^{a}$ can be. The reason is that the same averaged value will also enter into the threshold correction to the gauge kinetic function. 
From \eqref{eq:gauge_kin_fun_split}  (see also \eqref{eq:non_univ_threshold}), we see that the threshold correction has three contributions, each proportional to the flux integers in~\eqref{eq:flux_dependent_integers}. Let us focus on the part that contains the $B$-field, since it is the only one that can give a non-zero contribution on the orientifold odd cycles. Writing in terms of the position-dependent dilaton one can see that 
\begin{align}\label{eq:non_univ_threshold_odd_cycle}
    &\delta \alpha^{-1}_i|_{\rm odd} = \delta_i \int_D \text{Im}[S]\mathcal{F}_Y\wedge B_2 = \sum_a\delta \alpha^{-1}_i|_{\rm odd}^a\,,\\&
    \text{with} \,\,\delta \alpha^{-1}_i|_{\rm odd}^a = \delta_i\frac{b^ap_{Ya}}{g_s^{a}}\,.
\end{align}
In the last equality we have used that $p_{Ya}=\int_D \mathcal{F}_Y\wedge \omega_a$ is only non-zero when integrated over the odd 2-cycles, and hence the relevant averaged string coupling will also be $g_s^{a}$ as in Eq.~\eqref{eq:g_s_odd_cycle}. 
 Furthermore, from \eqref{eq:axion_linear_combs} we note that $p_{Ya} \neq 0$ along all 2-cycles whose associated axions $c^a$ enter the definition of the ALP $b$, and therefore also along the D1-brane instanton wrapping this combination of 2-cycles.
 This allows us to relate the D1-brane instanton action (around the $a$-th cycle) and the threshold corrections induced by the term proportional to the odd axion VEVs,
\begin{equation}\label{eq:relation_D1_inst_and_threshold}
S_{D1}^{a}=2\pi\frac{|\delta\alpha_i^{-1}|_{\rm odd}^{a}}{\delta_i\, p_{Ya}}\,.
\end{equation}
In particular, if $g_s^{a}\ll 1$, then not only does the D1-instanton action grow, but also the threshold correction to the gauge coupling is expected to be large. Conversely, when the threshold correction $\delta\alpha_i^{-1}|^{a}_{\rm odd}$ is small, so will be  $S_{D1}^{a}$ and the axion $c^a$ obtains a heavy mass. For concreteness we  specialise to cases with $h^{1,1}_-(X_3)=1$ in Fig. \ref{fig:ga/ma_vs_action_from_Wnp_new} and \ref{fig:ga/ma_vs_action_from_Knp_new}. As we show in the Appendix \ref{app:odd_cycles} this is indeed the most conservative case with respect to the lightness of the ALP $c^a$.

\subsection{Coupling-to-mass ratio for non-universal ALPs}\label{sec:g_a/m_a_non_universal_ALP}

\begin{figure}[t]
    \centering  \includegraphics[scale=0.7]{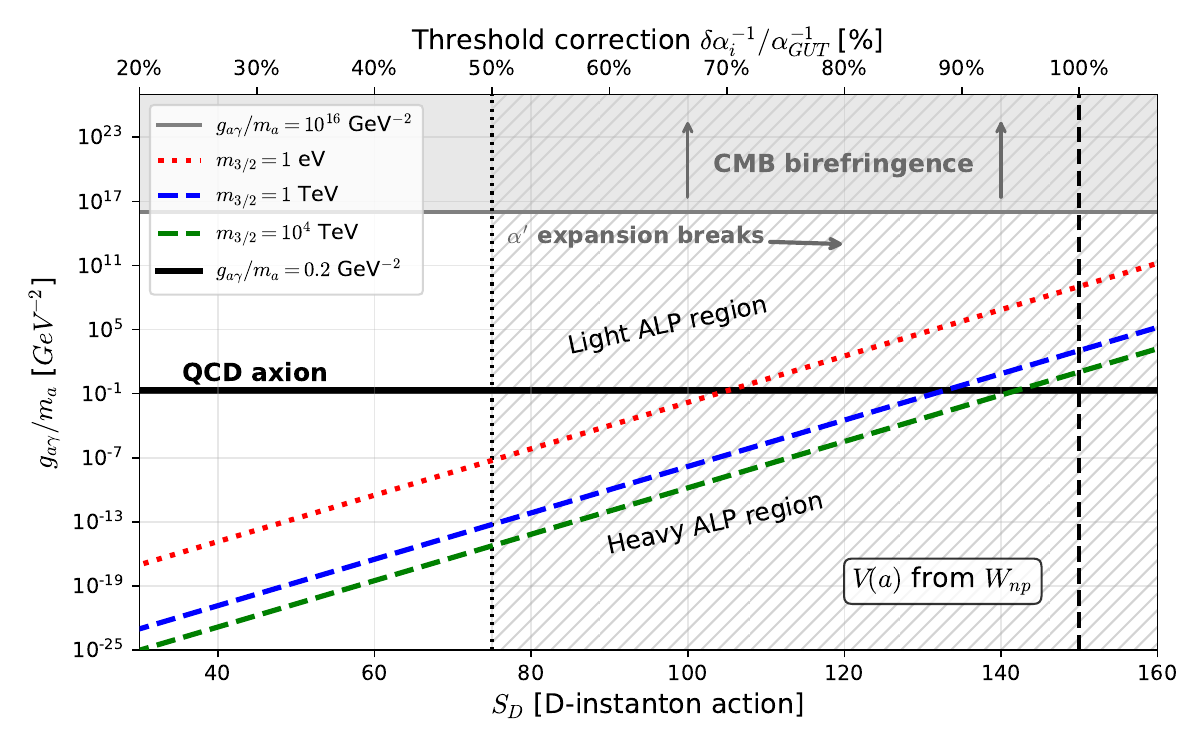}
    \caption{   
	Coupling-to-mass ratio for axions obtaining a mass from a \textbf{non-perturbative superpotential}, $W_{\rm np}$, as a function of the D-instanton action and the threshold correction, for a Pfaffian ${\cal A} = {\cal O}(1)$. We assume the MSSM value $\alpha_{\rm GUT}=1/24$ and comparable instanton actions, $S_{D(-1)}\sim S_{D1}$ (see text for details). In this case the UV scale is given by $\Lambda \sim (m_{3/2}M_{\rm GUT}^3)^{1/4}$, with different lines corresponding to different values for the gravitino mass. The dotted red line corresponds to $m_{3/2}=1 \text{ eV}$ and requires gauge mediated susy breaking with low-mass messenger, $M_{\rm mess}\lesssim 10^6$ GeV, which may be difficult to realize in models with fully stabilised moduli  
    (see text for details).
    Recall that high-scale threshold corrections larger than $\sim \mathcal{O}(\text{few})\%$, that is the precision of gauge coupling unification in the MSSM case, require additional charged matter at intermediate scales that also change the boundary condition for the unified gauge coupling, $\alpha_{\rm GUT}^{\rm new} > \alpha_{\rm GUT}=1/24$. This tends to decrease the UV instanton action, as discussed in Section~\ref{sec:large_threshold_and_ga/ma}. The gray region corresponds to a coupling-to-mass ratio that allows cosmic birefringence. Note that even if the threshold correction is as large as the tree-level contribution, axion-induced cosmic birefringence cannot be achieved. The hatched region corresponds to the instanton action values for which the $\alpha^\prime$ expansion breaks down (assuming $g_s^{a}\sim g_s^{D}$), see Section \ref{sec:alpha_prime_expansion} for details. In the hypothetical case of parametrically small Pfaffian, $\mathcal{A}\ll 10^{-66}$, the D$(-1)$/D1-instanton potential is negligible. An axion potential will be generated from a D3-brane induced superpotential with $S_{D3}\sim 2\pi/\alpha_{\rm GUT}$. This corresponds to the vertical dashed black line, see Section \ref{sec:Pfaffian_suppression} for more details.} 
	\label{fig:ga/ma_vs_action_from_Wnp_new}
\end{figure}

\begin{figure}[t]
    \centering  \includegraphics[scale=0.7]{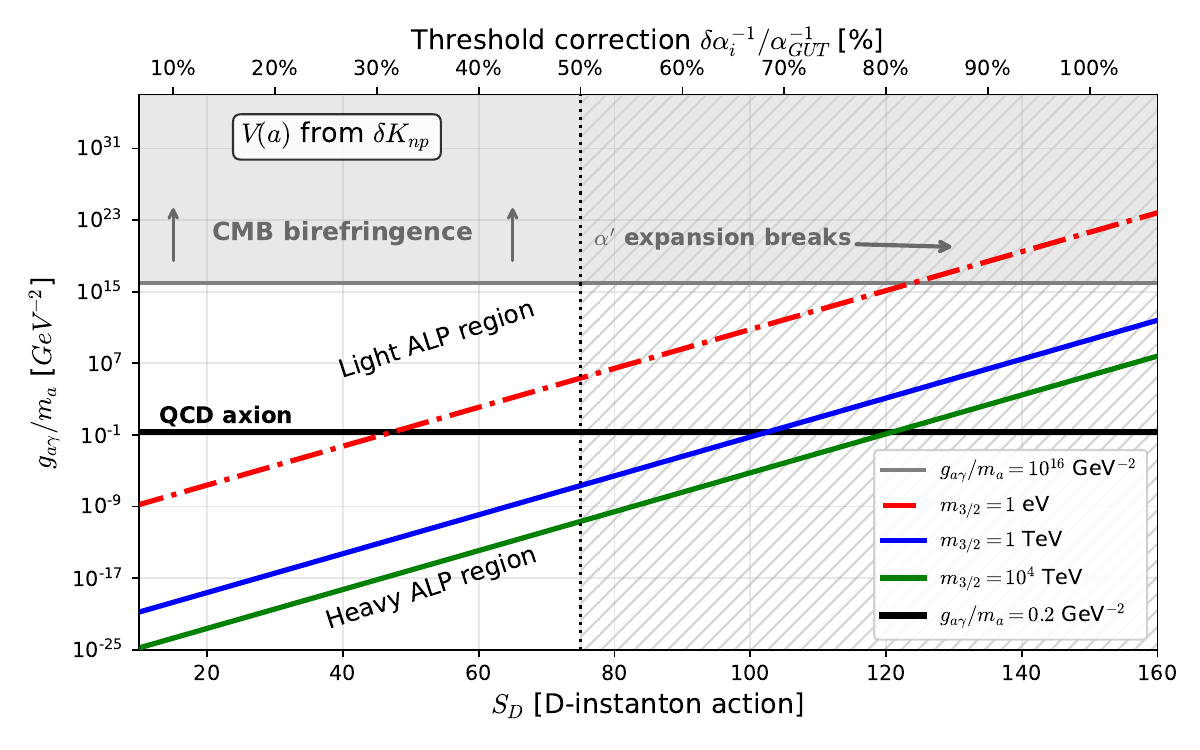}
    \caption{Coupling-to-mass ratio for axions obtaining a mass from a \textbf{non-perturbative K\"ahler potential}, $\delta K_{\rm np}$, as a function of the D-instanton action and the threshold correction, for a moduli-dependent instanton prefactor ${\cal A}_K = {\cal O}(1)$. We assume the usual MSSM value $\alpha_{\rm GUT}=1/24$ and comparable instanton actions, $S_{D(-1)}\sim S_{D1}$ (see text for details). In this case the UV scale is given by $\Lambda \sim (m_{3/2}M_{\rm GUT})^{1/2}$ and different lines correspond to different values of the gravitino mass. 
    Similar to Figure \ref{fig:ga/ma_vs_action_from_Wnp_new}, the same comments for the string coupling, $\alpha^\prime$ expansion, and large threshold corrections apply.  
    }	
	\label{fig:ga/ma_vs_action_from_Knp_new}
\end{figure}

From the discussion in the previous section, we conclude that the instanton potentials generated by D$(-1)$ and D1-instantons are unsuppressed and lead to heavy axions when threshold corrections are small, even when the variation of $g_s$ is taken into account. As a result, in F-theory GUTs, non-universally coupled ALPs, defined by the linear combination in Eq.~\eqref{eq:axion_linear_combs}, satisfy the relation~\eqref{eq:coupling-to-mass-ratio} due to their large masses. 

In this section we make this precise by estimating the ratio $g_{a\gamma}/m_a$ for the ALPs $C_0$ and $c^a$. Assuming $\mathcal{A}\sim\mathcal{O}(1)$, we have shown that the axion potentials $V_{D(-1)}$ and $V_{D1}$ are only partially screened by low-energy supersymmetry: For an axion potential that originates from a superpotential,  $\Lambda^4 = m_{3/2}M_{\rm GUT}^3$, see Eq.~\eqref{eq:Lambda4_superpotential}. On the other hand, as shown in Eq.~\eqref{eq:Lambda4_Kahler_potential}, when the potential comes from non-perturbative corrections to the K\"ahler potential, one pays an extra power of SUSY breaking suppression and the amplitude of the axion potential becomes $\Lambda^4= m_{3/2}^2M_{\rm GUT}^2$. 

By combining the expressions
\begin{equation}\label{eq:ma_and_ga_expressions}
    m_a^2 \sim \frac{\Lambda^4}{F_a^2} e^{-S_D}\,,\,\,\,\, \quad g_{a\gamma}\sim \frac{\alpha_{\rm em}}{2\pi F_a} \,,
\end{equation}
one sees that the ALP coupling-to-mass ratio is independent of $F_a$ and exponentially sensitive to the instanton action (due the dependence of the axion mass), 
\begin{equation}
    \frac{g_{a\gamma}}{m_a}\sim \frac{\alpha_{\rm em}}{2\pi}\frac{e^{S_D/2}}{\Lambda^2}\, .
\end{equation}
Here $S_D$ is either the D$(-1)$ or D1-instanton action, depending on whether we are estimating the ratio for the $C_0$ or $c^a$ axions.

In Fig.~\ref{fig:ga/ma_vs_action_from_Wnp_new} we show the prediction for the coupling-to-mass ratio for different values of $m_{3/2}$ in the case that the ALPs obtain a mass from a superpotential (assuming that the actions are comparable $S_{D(-1)}\sim S_{D1}$, see Eq.~\eqref{eq:odd_inst_action}). We find that even with low-scale supersymmetry breaking (for example, as it can occur in gauge-mediated models), corresponding to a gravitino mass around $m_{3/2}\sim \mathcal{O}(1)$ eV, the coupling-to-mass ratio is expected to be much smaller than the QCD axion prediction unless the threshold corrections are large, around $\sim $ 70 \% of the tree-level contribution. 

In Fig.~\ref{fig:ga/ma_vs_action_from_Knp_new} we show the predictions for the coupling-to-mass ratio for ALPs which receive a mass from non-perturbative corrections to the K\"ahler potential, again assuming $S_{D(-1)}\sim S_{D1}$. In this case the scale of the potential is given by $\Lambda^4\approx m_{3/2}^2M_{\rm GUT}^2$, leading to lighter axions in the cases where the gravitino is light. Despite this, we note that even for the lowest values of the gravitino mass, $m_{3/2}\sim \mathcal{O}(1)$ eV, the threshold correction required for $g_{a\gamma}/m_a$ to lie above the QCD axion prediction is $\gtrsim 30 \%$ of $\alpha_{\rm GUT}^{-1}$. This will require additional matter at intermediate scales with exotic quantum numbers (forming incomplete GUT multiplets) to fix the value of the gauge couplings, as we will discuss in Section \ref{sec:large_threshold_and_ga/ma}. We will see, however, that in such scenarios the UV value of the gauge couplings becomes larger, the action of the D-instanton effects becomes smaller, and the ALPs become heavier. As a result, no light ALP breaking the relation \eqref{eq:coupling-to-mass-ratio} is present in that case.

\subsubsection{$g_{a\gamma}/m_a$ and cosmic birefringence}\label{sec:DW_induced_birefringence}
It is interesting to compare the results for the coupling-to-mass ratio $g_{a\gamma}/m_a$ of $C_0$ and $c^a$ with the required values to generate cosmic birefringence, that is the rotation of the CMB photons polarization as they travel through the universe~\cite{Minami:2020odp,Eskilt:2022cff,Diego-Palazuelos:2023mpy,Galaverni:2023zhv,Diego-Palazuelos:2025dmh,Carralot:2026kps}. In the standard approach with a slow-rolling axion coupled to photons~\cite{Carroll:1998zi,Finelli:2008jv,Komatsu:2022nvu,Gasparotto:2023psh}, cosmic birefringence occurs for axions with mass lighter than the Hubble parameter at CMB, $m_a\lesssim H_{\rm CMB}\approx 10^{-28}$ eV. The field value changes between the CMB time and today, generating an isotropic rotation of the polarization of CMB photons. Assuming that the axion decay constant is bounded from above by the Planck scale, $F_a\lesssim  M_{\rm Pl}$, we find that to generate axion-induced cosmic birefringence one needs
\begin{equation}\label{eq:coupling-to-mass_birefringence}
    \frac{g_{a\gamma}}{m_a}\gtrsim 5\times 10^{15} \text{ GeV}^{-2}\,.
\end{equation}
This corresponds to the shaded region in light gray in Figures \ref{fig:ga/ma_vs_action_from_Wnp_new} and \ref{fig:ga/ma_vs_action_from_Knp_new}.

Cosmic birefringence with an axion heavier than $H_{\rm CMB}$ is also possible in the case of meta-stable axionic domain-wall networks that survive after CMB~\cite{Agrawal:2019lkr,Takahashi:2020tqv,Ferreira:2023jbu,Kaloper:2026ygk}.
If a photon traverses a wall once, its polarization rotates by an angle $\beta = \pm \frac{k_{\rm em }\alpha_{\rm em}}{N_{\rm DW}}$~\cite{Agrawal:2019lkr}, with $k_{\rm em }$ the anomaly coefficient with EM, and $N_{\rm DW}$ the domain wall number. 
A strong bound on the tension of these topological defects comes from measurements of the CMB. Imposing that the temperature fluctuations are small enough bounds the wall tension $N_{\rm DW}G_N\sigma/H_0\lesssim 5.6\times 10^{-6}$~\cite{Sousa:2015cqa}, with $\sigma \sim m_a F_a^2$ the wall tension and $G_N$ Newton's constant. This bound, together with the bound to the axion decay constant from astrophysics $F_a\gtrsim 10^8$ GeV, can be used to obtain a lower bound in the coupling-to-mass ratio for birefringence induced by axionic domain walls,
\begin{equation}
    \frac{g_{a\gamma}}{m_a}\gtrsim 10^{16}\text{ GeV}^{-2}\,,
\end{equation}
which is comparable to the case of a slow-rolling axion. This concludes that birefringence induced by vacuum interfaces is also incompatible with \eqref{eq:coupling-to-mass-ratio}.

Only in cases where the axion potential comes from the non-perturbative corrections to the K\"ahler potential  and the gravitino mass is $m_{3/2}\ll $ TeV, could one, at first sight, obtain light enough axions with a sufficiently large coupling-to-mass ratio. However, this would require non-GUT symmetric threshold corrections comparable to the tree-level contribution, and as noted already and discussed in more detail in Section \ref{sec:large_threshold_and_ga/ma}, 
 does not lead to violations of the bound \eqref{eq:coupling-to-mass-ratio}.

\subsection{Possible loopholes} \label{sec_loopholes}

Our analysis up to here has shown that coupling-to-mass ratios for ALPs above the QCD axion prediction, $\frac{g_{a\gamma}}{m_a}\sim \mathcal{O}(1)\frac{\alpha_{\EM}}{f_\pi m_\pi}$, are only possible if the D$(-1)$ or D1-instanton potentials are strongly suppressed. 
 There are two ways how this can happen -- either because of 
large  D$(-1)$/D1-instanton actions or because of strongly suppressed Pfaffian prefactors, which so far have been assumed to be $\mathcal{A}\sim{\cal O}(1)$.
  In this section we argue that even these scenarios do not lead to an arbitrarily strong violation of the bound  \eqref{eq:coupling-to-mass-ratio} within regions of control for the effective action.

\subsubsection{Large D-instanton action and $\alpha^\prime$ expansion}
\label{sec:alpha_prime_expansion}

The first potential loophole to the bound \eqref{eq:coupling-to-mass-ratio} is that the tree-level  D1- or D$(-1)$-instanton action becomes large.
 We have already observed in Sections \ref{sec:D-1_instantons} and \ref{sec:D1_instantons} that this requires large non GUT-symmetric threshold corrections. Taking into account the necessary additional matter to compensate for such splittings of the gauge couplings in fact leads to a weaker, rather than stronger, instanton suppression, as we will discuss in Section \ref{sec:large_threshold_and_ga/ma}.
 Independently of this line of reasoning, such large instanton actions cannot occur without leaving the regime of control for the effective action.
  Indeed, realistic F-theory GUTs require a region of the compact space to have $g_s\sim \mathcal{ O}(1)$, implying that $C_0$ always obtains a mass which is at least of the order of the gravitino mass, see Section \ref{sec:D-1_instantons}. 
On the other hand, one could wonder if the $c^a$ ALP can become light because $S_{D1}$ becomes large. If the axion potential comes from non-perturbative corrections to the K\"ahler potential and the gravitino is very light ($m_{3/2}\sim 1$ eV), a D1-instanton action larger than $\sim 30 \%$ of $\frac{2\pi}{\alpha_{\rm GUT}}$ (where $\alpha_{\rm GUT}=1/24$) allows a coupling-to-mass ratio for the ALP $c^a$ larger than the QCD axion, see Fig.~\ref{fig:ga/ma_vs_action_from_Knp_new}. As  $S_{D1}$ becomes larger, as a result of a smaller average string coupling $g_s^{a}$, both the coupling-to-mass ratio and the non-universal threshold correction in Eq.~\eqref{eq:non_univ_threshold_odd_cycle} grow. However, above a certain limit control of the theory is lost. The reason is that when $g_s^{a}$ is comparable to the averaged value over the GUT divisor, $g_s^{D}$, the $\alpha^\prime$ expansion ceases to be under control if $S_{D1}$ is large. This corresponds to the hatched region in Figures \ref{fig:ga/ma_vs_action_from_Wnp_new}, \ref{fig:ga/ma_vs_action_from_Knp_new} and will be discussed in more detail around Eqs.~\eqref{eq:D3_andD-1_actions} and  \eqref{eq:threshold_correction_varying_gs_and_SD1} in the context of specific examples; there we will show that as the D$(-1)$ or the D1-brane action increase, the D3-brane action decreases.

\subsubsection{Pfaffian suppression }\label{sec:Pfaffian_suppression}

The second potential loophole to the bound \eqref{eq:coupling-to-mass-ratio} is that the 
Pfaffian-like factor $\mathcal{A}$ or $\mathcal{A}_K$ in front of the superpotential or K\"ahler potential corrections of Section \ref{eq:V_D(-1)} and \ref{eq:V_D1}
 is very small. 

For a moderately small Pfaffian, $\mathcal{A}\lesssim O(1)$, our previous analysis applies and the non-universal ALPs obtain a large mass. As noted in  Footnote \ref{footnote:CSaxion_Pfaffian}, the assumption $\mathcal{A}\lesssim 1$ is consistent with the assumption of small quantum corrections from the complex structure moduli to the instanton action, and of absence of extra light fermion instanton modes.  Conversely, $\mathcal{A}$ decreases when vector-like fermionic modes become lighter and lighter in moduli space or when the complex structure moduli dependent quantum corrections to the instanton effective action become unsuppressed (and contribute in such a way that the total instanton action is increased); as a result, the D-instanton potential becomes shallower and the ALP mass smaller. For example, for $\mathcal{A}\lesssim  10^{-66}$, the axion mass from the D$(-1)$-instanton potential is smaller than the contribution from a D3-brane induced axion potential (with $S_{D3}\sim 2\pi/\alpha_{\rm GUT}$) even if $S_{D(-1)}\sim O(1)$. 
   It is expected that 
 a strong suppression of the Pfaffian as a result of large quantum corrections to the instanton action
 comes correlated with quantum corrections of similar magnitude for the gauge kinetic functions and is in tension with control of the effective field theory. Since the precise moduli dependence of the Pfaffian is not known in general, more quantitative statements are challenging to make in full generality. Nevertheless, we note that this extreme situation is likely in conflict with perturbative control as well as with obtaining the right gauge couplings in the IR.

As the Pfaffian decreases below this value, $\mathcal{A}\ll  10^{-66}$, the D$(-1)$/D1-instanton contributions to the ALP potential become negligible. However, as we will argue now, in this case other instanton contributions guarantee a non-vanishing axion potential. This is expected already from the point of view of the Axion Weak Gravity Conjecture \cite{Arkani-Hamed:2006emk}, which postulates that the instanton suppression factor for an axion with decay constant $F_a$ is bounded as $S_{\rm inst} \leq \alpha \frac{M_{\rm Pl}}{F_a}$ with $\alpha = {\cal O}(1)$. By interpreting the Pfaffian as the exponential of a loop correction to the instanton action, this general principle runs counter to a substantial cancellation of the axion potential by parametrically small Pfaffians, simultaneously for all instantons coupling to the axion. Indeed, in the case at hand the axion couples to gauge fields, and a non-perturbative potential is therefore guaranteed to be generated. Recall from \eqref{eq:axion_linear_combs} and \eqref{eq:axion_linear_combs-2} that there are two relevant linear combinations of axions, $\tilde a$ and the ALP $b$. Since the combination $\tilde a$ couples in a GUT symmetric way, it receives a mass from QCD effects. The ALP combination $b$, however, couples to hypercharge and $SU(2)$ gauge fields. The mass for the axion therefore has an irreducible potential given by
\begin{equation}\label{eq:D3_inst_potential_ALP}
    V(b) = -m_{3/2}M_{\rm GUT}^3\, e^{-S_{D3}}\cos (b)\,,
\end{equation}
with $S_{D3}=2\pi/\alpha_{w}$ or $S_{D3}=2\pi/\alpha_{Y}$, depending on which action is smaller. These effects come from a fluxed D3-brane instanton. Specifically, for the GUT breaking flux background \eqref{eq:defFflux}, an instanton along the GUT divisor $D$ with instanton flux ${\cal F}_S + \frac{1}{2} {\cal F}_Y$ behaves like an $SU(2)_w$ gauge instanton and consequently contributes to the superpotential (see \cite{Akerblom:2006hx} for an account of gauge instantons as D-brane instantons).\footnote{The instanton has charged zero modes. These can be absorbed via interactions in the instanton effective action paralleling the structure of the Yukawa couplings and couplings to gauge bosons in GUTs. See for example \cite{Demirtas:2021gsq} for an analogous discussion.}

 In Figure \ref{fig:ga/ma_vs_action_from_Wnp_new} we represent the action of this irreducible ALP potential~\eqref{eq:D3_inst_potential_ALP} by a vertical dashed black line. One can see that in this extreme case with $\mathcal{A}\ll 10^{-66}$,  for a gravitino mass above the TeV scale, $m_{3/2}\gtrsim \mathcal{O}(1)$ TeV, the ALP coupling-to-mass ratio is $g_{a\gamma}/m_a\lesssim 10^3$ GeV$^{-2}$, that is at most 3-4 orders of magnitude above the QCD axion prediction. While this situation allows for an ALP above the QCD axion line (with the concrete distance above the line determined by $m_{3/2}$, see Fig.~\ref{fig:ga_vs_ma_plot}), cosmic birefringence cannot be achieved; even if the gravitino is as light as $m_{3/2}\sim \mathcal{O}(1)$ eV, the predicted ALP coupling-to-mass ratio is around 8 orders of magnitude smaller than the minimal required for a birefringence signal, see Eq.~\eqref{eq:coupling-to-mass_birefringence}.

\section{Large threshold corrections and $g_{a\gamma}/m_a$}\label{sec:large_threshold_and_ga/ma}
 As shown in Fig.~\ref{fig:ga/ma_vs_action_from_Wnp_new} and \ref{fig:ga/ma_vs_action_from_Knp_new}, F-theory ALPs with $g_{a\gamma}/m_a$ larger than the QCD axion seem possible only for large D-instanton actions. Even for a very low gravitino mass $m_{3/2}\sim 1$ eV, a light ALP with $g_{a\gamma}/m_a$ larger than the QCD axion would require that the threshold correction must be above $\sim 70\%$ ($\sim 30$ \%) of the tree-level contribution when the axion potential comes from corrections to the  superpotential  (K\"ahler potential). For a TeV scale gravitino -- the minimal expected in gravity mediation -- the size of the required threshold corrections for an ALP above the QCD line change to $\sim 90\%$ in the case of an axion potential from $W_{\rm np}$ and $\sim 70$ \% when it comes from corrections to the K\"ahler potential. This leads, in general, to large non-universal corrections to the gauge couplings that spoil the successful prediction of gauge coupling unification. For example, for a fixed particle content with 3 MSSM families and no other degree of freedom up to the GUT scale, $\mathcal{O}(10)\%$ non-universal threshold corrections would lead to predictions for the gauge couplings incompatible with IR measurements. Phenomenologically viable models with large holomorphic threshold corrections require new, light degrees of freedom which do not come in complete GUT representations. These new representations beyond the 3 SM-like generations are required to modify the beta functions in a way that, for given UV values of the gauge couplings, the measured IR values for the gauge couplings can be recovered.\footnote{For this reason, in some sense the contributions of the new, incomplete multiplets pose a tuning problem.} 

In this section we study  the implications of these new states for the ALP coupling-to-mass ratio, $g_{a\gamma}/m_a$. A generic consequence of adding representations beyond the MSSM is that the UV value of the gauge couplings, which we call $\alpha_{i}(M_{\rm GUT})$, increase with respect to the standard (MSSM-like) unified prediction, which from now on we call $\alpha_{\rm GUT}^{\rm mssm}\sim 1/24$.
Therefore the UV instanton (Euclidean D-brane) action is reduced with respect to the standard predictions. As we will see, also discussed in \cite{Reig:2025dqb,Agrawal:2025rbr}, this reduces the coupling-to-mass ratio $g_{a\gamma}/m_a$ for the ALPs in such a way that no ALP above the QCD axion prediction appears. 

Increasing $\alpha_{i}(M_{\rm GUT})$ may also spoil the quality of the QCD axion and re-introduce the strong CP problem since the minima of potential generated by QCD and other UV instantons are not necessarily aligned. The details depend on which representations are added. 
For the sake of generality, we will first consider the general case and then consider situations where the IR values of the gauge couplings are recovered by adding Higgs color triplets with mass $M_{3}\ll M_{\rm GUT}$.\footnote{For simplicity we assume that the GUT group is broken to $SU(3)_C\times SU(2)_w\times U(1)_Y$, but our discussion holds in other cases, too.}

\subsection{Unification by incomplete multiplets at intermediate scales }\label{subsec:incomplete_multiplets}

Independently of their value at the GUT scale, $M_{\rm GUT} \sim M_s$, and independently of the value of $p_{YY},p_{YS},p_{Ya}$, the gauge couplings including the holomorphic threshold corrections in Eq.~\eqref{eq:generic_gauge_kin_function_corrected}, \eqref{eq:non_univ_threshold} satisfy the relation \cite{Blumenhagen:2008aw}
\begin{equation}\label{eq:coupling_condition_Ms}
    \frac{1}{\alpha_Y(M_{\rm GUT})} = \frac{1}{\alpha_w(M_{\rm GUT})} + \frac{2/3}{\alpha_s(M_{\rm GUT})}\,.
\end{equation}
In the case $\alpha_s(M_{\rm GUT}) = \alpha_w(M_{\rm GUT})$ this relation leads to the standard GUT prediction. 

The corrections in Eq. \eqref{eq:non_univ_threshold} shift the gauge couplings in a non-universal way. For negative $p_{YY}$ (as expected for supersymmetric compactifications \cite{Blumenhagen:2008zz}), the threshold corrections induce the following ordering of gauge couplings: 
\begin{equation}\label{eq:ordering_gauge_coupling_at_Ms}
    \frac{1}{\alpha_s (M_{\rm GUT})} < \frac{3}{5}\frac{1}{\alpha_Y (M_{\rm GUT})} < \frac{1}{\alpha_w (M_{\rm GUT})}\,.
\end{equation}
This relation suggests that, in scenarios with incomplete GUT multiplets, the change to the beta function of QCD is expected to be larger than to hypercharge, which is in turn expected to be larger than the change to the weak $SU(2)$. 
This relation differs from standard GUT relations and can be easily recovered from the imaginary part of the (threshold corrected) gauge kinetic function in Eq.~\eqref{eq:gauge_kin_fun_split} (see also \eqref{eq:non_univ_threshold}). A given theory is phenomenologically viable if the gauge couplings satisfy~\eqref{eq:coupling_condition_Ms} and Eq.~\eqref{eq:ordering_gauge_coupling_at_Ms}  at the GUT scale, and has the right matter content with masses such that these values are RGE evolved to the measured values in the IR.

We proceed now to relate the string coupling, and therefore the relevant D-instanton actions, to the non-universal running of each SM gauge factor. To fix ideas, we will consider first a case where $g_s$ is approximately constant, postponing the discussion of varying string coupling to the end of the section.  To find the relation between $g_s$ and $S_D$ we first estimate how the new, incomplete multiplets modify the value of the gauge coupling at the scale $M_{\rm GUT}$ with respect to the standard MSSM prediction, $\alpha_{\rm GUT}^{\rm mssm}=1/24$. From a (bottom-up) 4d EFT perspective, that is by taking the observed values of the gauge couplings at low-energy and evolving them towards the UV, the corrected value of the gauge coupling at the GUT scale can be written as
\begin{equation}\label{eq:new_GUT_coupling}
    \frac{1}{\alpha_{i}(M_{\rm GUT})} = \frac{1}{\alpha_{\rm GUT}^{\rm mssm}} + \Delta \alpha^{-1}_i\,,\,\,\text{ with }\,\,\Delta\alpha^{-1}_i = \frac{\Delta \beta_i}{2\pi}\log \left ( \frac{M_{\rm GUT}}{M_{\rm new}}\right )\,.
\end{equation}
Here $\Delta \beta_i$ denotes the change to the 1-loop beta function of the $i$-th gauge factor induced by incomplete multiplets, with mass $M_{\rm new}< M_{\rm GUT}$. The $\Delta\alpha_i^{-1}$ are expected to be different for each group and also smaller than $(\alpha_{\rm GUT}^{\rm mssm})^{-1}$ to ensure both perturbativity and approximate unification of couplings. As expected when  new charged degrees of freedom are added, the shift $\Delta \alpha_i^{-1}$ is negative, which implies that $\alpha_i(M_{\rm GUT})>\alpha_{\rm GUT}^{\rm mssm}$.
Combining these equations with the definition of the gauge kinetic function in \eqref{eq:generic_gauge_kin_function_corrected} (see also \eqref{eq:non_univ_threshold}), we obtain
\begin{equation}\label{eq:general_D-1_action_as_function_of_shifts}
    \frac{2\pi}{g_s} = \frac{-2}{2p_{YS}+p_{YY}-2b^ap_{Ya}}2\pi\left ( \Delta \alpha^{-1}_w - \Delta \alpha^{-1}_s \right )\,.
\end{equation}
This equation relates, in a general way, the non-universal shifts $\Delta \alpha_i^{-1}$ for $SU(2)_w$ and $SU(3)_C$ (induced by new, incomplete multiplets) to the D$(-1)$-instanton action, $S_{D(-1)}=\frac{2\pi}{g_s}$. 

A few comments are in order. Assuming no cancellation between the integers, $2p_{YS}+p_{YY}-2b^ap_{Ya}\neq 0$, the first fraction is expected to be an $\mathcal{O}(1)$ number. In the limit $\Delta \alpha_w^{-1}=\Delta \alpha_s^{-1}$, as one expects when the new charged matter comes in full GUT multiplets, then  the instanton action vanishes $S_{D(-1)}\rightarrow 0$. This occurs because in that case unification requires small holomorphic threshold corrections (see \eqref{eq:gauge_kin_fun_split} and \eqref{eq:non_univ_threshold}). For generic values of $p_{YS},\, p_{YY} ,\, p_{Ya}$, the holomorphic threshold correction vanishes for large string coupling $g_s$ and in this case the D-instanton action also vanishes.  
Finally, assuming for simplicity that the fraction on the righthand side is positive, we see that the D$(-1)$-action is largest when there are no intermediate scale states with $SU(2)_w$ charge, that is when $\Delta \alpha_w^{-1}=0$.\footnote{If the fraction is negative, the instanton action is maximized for  $\Delta \alpha_s^{-1}=0$.} In the next section we will consider this possibility explicitly.

\subsection{Unification by Higgs triplets}

Let us consider, following \cite{Blumenhagen:2008aw}, an example where unification is achieved by light Higgs color triplets transforming as $H_3^i\sim (\mathbf{3},\mathbf{1},-1/3)$ under the SM, that is the color triplet partner of a weak doublet in an $SU(5)$ fundamental.\footnote{We note that, in the absence of symmetries or some other mechanism, this representation generates dimension-6 operators that lead to fast proton decay unless its mass is close to $M_{\rm GUT}$. This is the essence of the doublet-triplet splitting problem. However, for the sake of simplicity, we will assume that it can be much lighter than the GUT scale without specifying any particular mechanism.} Studying the running of the QCD gauge coupling in this example allows us to relate the mass of the ALP and the Higgs triplet. We will also obtain a relation between the action of the different D-instantons. For simplicity, we are assuming  $h^{1,1}_-(X_3)=0$, taking into account the general case later. 

Following the reasoning in Section~\ref{subsec:incomplete_multiplets}, we use the 1-loop evolution of the QCD gauge coupling and $p_{YY}\equiv\int_D  \mathcal{F}_Y\wedge \mathcal{F}_Y=-2$ (a value that avoids light exotics~\cite{Donagi:2008kj}), to obtain 
\begin{equation}\label{eq:gs_vs_change_gauge_coupling}
    \frac{1}{g_s}=-\frac{\Delta\beta_{\rm qcd}}{2\pi}\log \left ( \frac{M_{\rm GUT}}{M_{3}}\right )\,.
\end{equation}
Here $M_{3}$ is the mass of the color triplet, $M_{\rm GUT}\sim M_s$ is the unification scale, and $\Delta \beta_{\rm qcd} = -N_3$ is the change to the QCD one-loop beta function due to $N_3$ color triplet superfields. For $N_3 = 1$ this implies $M_{3}\approx M_se^{-2\pi/g_s}$, which, as noted in~\cite{Blumenhagen:2008aw}, resembles a situation where the Higgs triplet gets a mass from D$(-1)$-instantons.

An interesting feature of Eq.~\eqref{eq:gs_vs_change_gauge_coupling} is that it implies a simple relation between the D$(-1)$-brane and D3-brane instanton actions. First we note that in this case
\begin{equation}
    \Delta\alpha^{-1}_{s} = \frac{\Delta \beta_{\rm qcd}}{2\pi}\log \left ( \frac{M_{\rm GUT}}{M_{3}}\right ) = - \frac{S_{D(-1)}}{2\pi}\,.
\end{equation}
Consider now the action of the UV QCD gauge instanton -- that is, a QCD instanton of size $\rho\sim M_{\rm GUT}^{-1}$. This action depends on the new value of the QCD gauge coupling at $M_{\rm GUT}$ and coincides with the action of an Euclidean D3-brane wrapping $D$, given by 
\begin{equation}\label{eq:D3_andD-1_actions}
    S_{D3} = \frac{2\pi}{\alpha_{s}(M_{\rm GUT})} = \frac{2\pi}{\alpha_{\rm GUT}^{\rm{mssm}}} - S_{D(-1)}\,.
\end{equation}

This is an interesting relation between the action of the relevant D-instantons and the unified gauge coupling of minimal GUTs, $\alpha_{\rm GUT}^{\rm{mssm}}= 1/24$. It shows that as $S_{D(-1)}$ grows to make the $C_0$ ALP lighter (see Figs.~\ref{fig:ga/ma_vs_action_from_Wnp_new} and~\ref{fig:ga/ma_vs_action_from_Knp_new}), $S_{D3}$ decreases accordingly. A small D3-instanton action for QCD will in general (in the presence of generic CPV phases) introduce a quality problem for the would-be QCD axion, $c_D=\int C_4$. In some cases the UV instanton potential may surpass the IR contribution. {If the D3-instanton overcomes the QCD instanton potential, the $c_D$ axion becomes much heavier than the usual QCD axion mass relation and can be integrated out. In this case, due to its coupling to gluons (see Eq.~\eqref{eq:axion_linear_combs}) the $C_0$ axion obtains a potential from QCD. Furthermore, the $C_0$ potential contribution from D($-1$)-branes is too large for $C_0$ to solve the strong CP problem.} This implies that in the case of F-theory GUTs with large threshold corrections and Higgs triplets (assuming nearly constant $g_s$ and $h^{1,1}_- (X_3)=0$), {ALPs with a large coupling-to-mass ratio $g_{a\gamma}/m_a$ are not possible.}
Similar conclusions are expected to apply to scenarios with other representations following the more general reasoning in Section~\ref{subsec:incomplete_multiplets}, as we will see below. We emphasise that the case with only Higgs triplets allows for the lightest possible ALPs, as $\Delta \alpha_w^{-1}=0$ maximises $S_{D(-1)}$ (see Eq.~\eqref{eq:general_D-1_action_as_function_of_shifts}).

\subsubsection*{Threshold corrections in scenarios with odd cycles and varying $g_s$}
Let us now consider a case with orientifold-odd cycles, $h^{1,1}_-(X_3)\neq 0$, and varying dilaton. In this case we can re-write the above relation for $\Delta\alpha_i^{-1}$ and relevant instanton actions, see Eq.~\eqref{eq:general_D-1_action_as_function_of_shifts}, as 
\begin{equation}\label{eq:threshold_correction_varying_gs_and_SD1}
    2\pi(\Delta\alpha_w^{-1}-\Delta\alpha_s^{-1}) = -\frac{2\pi}{g_s^{D}}\left ( \frac{2p_{YS}+p_{YY}}{2}\right )+p_{Ya}S_{D1}^a\,,
\end{equation}
with $g_s^{D}$ being the average over the GUT divisor.
We have also used $S^a_{D1}=2\pi b^a/g_s^{a}$, with $g_s^{a}$ the averaged string coupling over the $a$-th odd 2-cycle.

To put some numbers to this, we consider the case where there is a single axion associated to an odd 2-cycle, $h^{1,1}_-(X_3)=1$.\footnote{This case is not only simpler, it is also the most conservative situation: it corresponds to the case that allows for the lightest ALPs from $C_2$, as shown in Appendix~\ref{app:odd_cycles}.} Let us also assume as before that $\Delta\alpha_w^{-1}=0$, and that the threshold correction associated with $C_0$ is negligible -- that is, $g_s^{D}\gg g_s^{a}$, so that $\Delta\alpha_s^{-1}\approx -\frac{p_{Ya}S_{D1}}{2\pi}$. These assumptions are made to study the case that allows for the lightest $C_2$ axion. To show what goes wrong when we try to make the ALP $c^a$ very light, we assume an instanton action $S_{D1}$ which is around 40 \% of the SM value, $S_{D1} \sim 0.4 \times 2 \pi/\alpha_{\rm GUT}^{\rm mssm}$. This implies that the threshold corrections and $\Delta\alpha_i^{-1}$ are also around 40\% of $(\alpha_{\rm GUT}^{\rm mssm})^{-1}$, and suggests that the SM gauge coupling at the compactification scale is around $\alpha_s(M_{\rm GUT})=\alpha_{\rm GUT}\sim 1/14$, see eq.~\eqref{eq:new_GUT_coupling}. Euclidean D3-branes wrapping the GUT divisor then generate a large UV instanton potential for the axion $c_D$, 
\begin{equation} \label{eq:VD3}
    V_{D3}(c_D) = -m_{3/2}M_{\rm GUT}^3e^{-S_{D3}}\cos (c_D+\delta_{D3})\, ,
\end{equation}
with $\delta_{D3}$ being a CP violating phase that depends on UV physics. It is easy to check that for a gravitino mass above $m_{3/2}\gtrsim \mathcal{O}(1)$ eV (TeV), the contribution to the axion mass from $V_{D3}(c_D)$ surpasses the IR QCD contribution as long as $\alpha_s(M_{\rm GUT})=\alpha_{\rm GUT}\gtrsim 1/18$ (1/22). In the case at hand, $m_{3/2}=1$ eV and $\alpha_s(M_{\rm GUT})=\alpha_{\rm GUT}\sim 1/14$, this potential generates a mass for $c_D$ which is $\sim 10^4$ times larger than the (IR) QCD potential contribution, introducing a severe quality problem for generic values of $\delta_{D3}$. See \cite{Benabou:2026jtv} for a related discussion in the context of the heterotic string.

The ALP $c^a$ can be made light in this setup, but at the cost of making $c_D$ very heavy due to the potential from D3-branes. As $C_0$ is always at least as heavy as the gravitino (due to effects of varying $g_s$) we can safely integrate it out. This means that either $c^a$ or $c_D$ -- but, crucially, not both -- can remain as a light axion in the theory. In the case where $c^a$ is light, $c_D$ is heavy and can be integrated out; as a result $c^a$ is now the axion coupling to QCD, and hence its coupling-to-mass ratio lies on the QCD line. Note, however, that the D1-instanton contribution (Eq.~\eqref{eq:V_D1}) is still too large for $c^a$ to solve the CP problem, as we expect that the D1-instantons will generically be misaligned with respect to the QCD instantons, so this case reintroduces the strong CP problem unless there is an underlying reason for the phases to be aligned. We remark that in cases with $h^{1,1}_-(X_3)>1$ the axions $c^a$ are not expected to be as light as in the $h^{1,1}_-(X_3)=1$ case, as discussed in more detail in App.~\ref{app:odd_cycles}. All in all, we find that the bound in Eq.~\eqref{eq:coupling-to-mass-ratio} continues to be satisfied in models with orientifold-odd 2-cycles, varying $g_s$, and large threshold corrections.

\section{Open string axions in F-theory GUTs}\label{sec:open_string_axion}

F-theory GUTs with intersecting 7-branes involve chiral matter at the intersection of branes, which defines a 6d worldvolume. In some cases, this leads to the appearance of axions from the phases of complex scalars, $\Phi=|\Phi|e^{i\varphi}$, on the matter curve on the compactification space where the 4-cycles wrapped by the 7-branes intersect. Such axion $\varphi$ is referred to as an open string axion in the D-brane literature \cite{Cicoli:2012sz,Cicoli:2013cha,Cicoli:2013ana,Cicoli:2022fzy,Petrossian-Byrne:2025mto,Loladze:2025uvf}. In this section we point out that open string axions also obey the bound \eqref{eq:coupling-to-mass-ratio}.

Suppose that the theory contains, in addition to the original GUT group, a $U(1)_C$ gauge factor along with a GUT singlet $\Phi$ charged under $U(1)_C$. The latter arises from a matter curve not contained in the GUT divisor. Depending on the precise $U(1)_C$ charged chiral matter, there may arise cubic anomalies of the type $[U(1)_C]^3$, $[G_{\rm GUT}]^2\times U(1)_C$ or $[G_{i}]^2\times U(1)_C$, with $G_i$ a SM group factor. The latter non-GUT symmetric anomalies are only relevant in cases where the gauge symmetry is broken to the SM via fluxes such that there are chiral superfields transforming in incomplete GUT multiplets. The discussion is similar to the case of boundary-localised matter in orbifold GUTs~\cite{Agrawal:2025rbr}. 

Via a generalised Green-Schwarz mechanism, the RR fields $C_{0},\,C_{2},\,C_{4}$ can cancel the cubic anomalies of the $U(1)_C$ gauge symmetry if they transform appropriately~\cite{Aldazabal:2000dg}.
The relevant part of the action is
\begin{equation}
    S_{\rm tot}=S_{\rm CS}+S_{\rm inter.}\,,
\end{equation}
where $S_{\rm CS}$ corresponds to the CS part of the $7$-brane action in Eqs.~\eqref{eq:CS_C_4} and~\eqref{eq:non_univ_correction_CS_action}, and $S_{\rm inter.}$ is the part of the action that contains the chiral superfields at the intersection. The action $S_{\rm tot}$ is gauge invariant provided that the RR fields shift under the localized (pseudo-anomalous) $U(1)_C$ gauge symmetry as 
\begin{equation}
    C_4 \rightarrow C_4  + q_C^{(4)}\delta(D) \lambda_C \,,\,\,C_2 \rightarrow C_2 + q_C^{(2)}\delta(\Sigma^-) \lambda_C \,,\,\,C_0 \rightarrow C_0 + q_C^{(0)}\lambda_C \,.
\end{equation}
The different integers $q_C^{(p)}$ determine which RR fields participate in the anomaly cancellation mechanism. In the equation above, $\delta (\Pi_p)$ is a generalized delta function acting as a $p$-form and $\lambda_C$ is the parameter of the gauge transformation. In practice some of the integers $q_C^{(p)}$ may be zero. For example, in cases where $q_C^{(2)}=q_C^{(0)}=0$, only $C_4$ shifts under $U(1)_C$. Since this field couples to the entire set of GUT gauge bosons as in Eq.~\eqref{eq:CS_C_4}, only GUT-symmetric anomalies of the type $[G_{\rm GUT}]^2\times U(1)_C$ can be cancelled. When the localised anomaly has a non-GUT symmetric form, the shifts of $C_2$ and $C_0$ can be used to cancel it.

The presence of this gauge transformation for the RR fields induces a gauge charge inflow from the bulk to the intersection that cancels the $U(1)_C$ anomalies. As recently discussed in \cite{Loladze:2025uvf}, for $|\Phi| \neq 0$, corresponding to a Higgsing of the pseudo-anomalous $U(1)_C$,this results in a mixing of the phase $\varphi$ and the axions coming from $C_4$, $C_2$, and $C_0$.\footnote{This is in one-to-one correspondence with field theoretic axions (phases of complex scalars) in the heterotic string, where they mix with the higher-form axions from $B_2$ and $B_6$ in scenarios with pseudo-anomalous $U(1)$s.} One linear combination -- the one that shifts under the gauge symmetry -- is eaten by the $U(1)_C$ gauge boson which gains a mass of the order of the string scale, $M_C^2\sim \alpha_C(F_a^2+|\Phi|^2)$, with $\alpha_C$ the $U(1)_C$ gauge coupling, and $F_a$ the closed string axion decay constant. The linear combinations which do not shift under the gauge symmetry remain in the 4d EFT as a perturbatively massless axions. The low-energy couplings to gauge bosons for these un-eaten axions are uniquely determined by the anomaly structure. Equivalently, the couplings to gauge bosons of the uneaten combinations are inherited from the CS couplings of the original RR fields to the gauge fields in the worldvolume of the 7-brane. 

The un-eaten axions satisfy the bound~\eqref{eq:coupling-to-mass-ratio} provided that the RR axions in the equivalent theory without open string axions satisfy it. This occurs because the same non-perturbative effects that break the shift symmetry of the RR axions generate a potential for the axions which survive at low energies. The potential in theories with anomalous U(1)‘s goes like $V\sim - |\Phi| \Lambda^3e^{-S_D}\cos (\theta)$, where $\theta $ is the gauge-invariant combination of axions, and $\Lambda, S_D$ are the overall scale and instanton action that would contribute to the potential in an anomaly-free theory with only closed string axions. In other words, the un-eaten axion $\theta$ (composed of RR axions and $\varphi$) inherits the quality of the higher-form axion~\cite{Loladze:2025uvf}. Since the couplings to gauge bosons are interited from the CS couplings of the RR fields, $c_D$, $C_0$, and $c^a$, and the shift symmetry has the same quality, the un-eaten axions will also satisfy the bound~\eqref{eq:coupling-to-mass-ratio}.

The main difference with respect to the theory with axions arising from the closed string spectrum is related to the effective decay constant of the un-eaten axion, which in some cases can be smaller than the string scale. To illustrate this, let us consider a simple situation where the uneaten linear combination is composed of a single closed string axion $a$ and the phase of a complex scalar, $\Phi$.
The effective decay constant for the uneaten axion is given by
\begin{equation}
    \frac{1}{F_{\rm eff}^2} = \frac{1}{F_{ a}^2}+\frac{1}{|\Phi|^2}\,.
\end{equation}
If the VEV $|\Phi|\ll M_s$, this contribution dominates and we obtain $ F_{\rm eff}\sim |\Phi|$. In practice, however, this only occurs if the (moduli dependent) Fayet-Iliopoulos (FI) term contribution to the D-term potential that fixes the VEV is tuned to be small. 

Note that in the language of the elliptic four-fold associated with the 7-brane configuration in F-theory, 
giving a VEV $|\Phi| \neq 0$ corresponds to a complex structure deformation of the elliptic fibration which smoothens out the degeneration of the elliptic fiber over the matter curve on which the charged field $\Phi$ is localised.
What in Type IIB language describes the resulting mixing between the RR and the open string axions therefore 
corresponds to a mixing of RR axions with the axionic components of the complex structure moduli of the four-fold.
Indeed, by starting from the singular locus in complex structure moduli space where $| \Phi|=0$, this illustrates that in general the RR axions mix with the complex structure moduli of the fourfold.
In the interior of moduli space, i.e. for large $|\Phi|$, this effect is in general challenging to quantify. However, the Type IIB analysis shows that the derived bounds for the pure RR axions are respected through the smoothing process. This, in turn, justifies focusing on the RR axions in our analysis.

\section{Implications for phenomenology}\label{sec:pheno}
We have provided evidence that the axion parameter space depicted in Figure \ref{fig:ga_vs_ma_plot}, especially the axion-photon coupling, is very restricted in F-theory GUTs, where every axion satisfies \eqref{eq:coupling-to-mass-ratio}  (modulo the potential loophole discussed in Section \ref{sec:Pfaffian_suppression}). The discovery of an ALP violating this bound -- that is, whose coupling-to-mass ratio lies above the QCD axion line -- would be incompatible with F-theory GUTs. 

In this direction, cosmic birefringence is the most dramatic example as it requires axion masses comparable to the current Hubble scale, $H_0\lesssim m_a \lesssim H_{\rm CMB}$, with photon couplings $g_{a\gamma} \sim \mathcal{O}(1) /f_a$. 
Remarkably, there is currently a hint for such a signal in the CMB that originates from analyses of WMAP, Planck, and ACT polarization~\cite{Minami:2020odp,Eskilt:2022cff,Diego-Palazuelos:2023mpy,Galaverni:2023zhv,Diego-Palazuelos:2025dmh,Carralot:2026kps}. 
These point towards a non-zero rotation angle $\beta \sim 0.3^{\circ}$ at around $3\sigma$ statistical significance. 
The birefringence hint will be tested by the next generation of CMB polarization experiments.  Thanks to new methods to improve the absolute polarization calibration, the Simons Observatory \cite{SimonsObservatory:2018koc,2023RScI...94l4502M} (already coming online) is expected to mitigate some of the dominant systematic effects affecting the current Planck-based analyses. A definitive confirmation of the signal above the 5$\sigma$ level, however, may require  LiteBIRD~\cite{LiteBIRD:2022cnt}.
Several axion-based explanations have been proposed, ranging from axion dark energy~\cite{Carroll:1998zi,Finelli:2008jv,Komatsu:2022nvu,Gasparotto:2023psh}, an axion string network~\cite{Agrawal:2019lkr,Jain:2021shf,Jain:2022jrp,Yin:2023vit}, and vacuum interfaces or domain walls~\cite{Takahashi:2020tqv,Ferreira:2023jbu,Kaloper:2026ygk}. The common ingredient in these explanations is an axion with a coupling-to-mass ratio $g_{a\gamma}/m_a\gtrsim 5\times 10^{15} \text{ GeV}^{-2}$, more than 16 orders of magnitude above the QCD axion prediction, see the discussion in Section \ref{sec:DW_induced_birefringence}.

Another salient feature of the class of F-theory GUT models studied in this work is that the QCD axion mass is expected to lie around the neV scale. As explained in Section~\ref{sec:dec_const_and_QCD_axion}, in minimal models the QCD axion arises from closed string modes. As the $C_0$ and $c^a$ axions are very heavy when the threshold corrections are small, the QCD axion linear combination comes almost entirely from the RR $C_4$ field, with negligible effects from axion mixing. This is in analogy to the QCD axion being given by the model-independent axion in weakly curved heterotic compactifications~\cite{Benabou:2026jtv}. As discussed around Eq.~\eqref{eq:expected_Fa}, this points to a QCD axion  mass around
\begin{equation}
    m_a^{\rm qcd}\approx 0.5\text{ neV}\,,
\end{equation}
which motivates some experiments such as the ABRACADABRA experiment~\cite{Kahn:2016aff,Ouellet:2018beu,Ouellet:2019tlz,Salemi:2021gck} (together with the follow-up DMRadio program~\cite{DMRadio:2022pkf,DMRadio:2022jfv,Benabou:2022qpv,DMRadio:2023igr,Ankel:2026zrv}), and related ideas using superconducting resonant frequency cavity conversion~\cite{Berlin:2019ahk,Giaccone:2022pke}. Other experiments able to access this mass range include axion-induced spin-precession experiments such as CASPEr~\cite{Graham:2013gfa,Budker:2013hfa,JacksonKimball:2017elr,Aybas:2021cdk,Dror:2022xpi}. Open string axions -- that is, axions from phases of complex scalar fields -- could in principle modify this expectation, but in generic situations the complex scalar VEVs lie near the string scale. This is because the moduli dependent FI terms tend to bring $|\Phi|$ close to $M_s$. See Section \ref{sec:open_string_axion} for a more detailed discussion.

We have considered F-theory GUT compactifications with orientifold-odd 2-cycles, whose number is given by $h^{1,1}_-(X_3)$. Therefore, in addition to the $h_+^{1,1}$ axions from integrating $C_4$ over 4-cycles (one of which corresponds to the GUT divisor), there exist $h^{1,1}_-(X_3)$ ALPs from $C_2$ integrated over the odd cycles, and the RR axion $C_0$. These fields couple via the holomorphic threshold corrections discussed throughout the paper, highlighting the importance of considering these corrections when studying the phenomenology of the axiverse. The $h^{1,1}_+(X_3)-1$ axions (subtracting the one that couples to the GUT divisor via the CS coupling \eqref{eq:CS_C_4}) arising from integrating $C_4$ over the other 4-cycles in the manifold are expected to have suppressed interactions with the Standard Model sector, since their couplings only appear via the kinetic and mass mixing with the QCD axion. This is equivalent to the phenomenon of kinetic isolation discussed in \cite{Gendler:2023kjt}, which would be interesting to verify in F-theory GUTs. All together, modulo the complex structure axions which are in general expected to be heavy (see Section \ref{sec:complex_structure_axions} for details), the total number of closed string axions is $h^{1,1}_-(X_3)+h^{1,1}_+(X_3)+1$, as expected.

\subsubsection*{ALP mass from superpotential}
The masses of ALPs other than the QCD axion can span many orders of magnitude depending on the size of the threshold corrections, the gravitino mass, and whether their potential arises from non-perturbative corrections to the superpotential or the K\"ahler potential. This is shown in Fig.~\ref{fig:ga_vs_ma_plot} for different benchmark points.

As an illustrative example, let us take a very simplified picture of the F-theory GUT axiverse where stringy thresholds are small, $\delta\alpha^{-1}_i\sim O(10)\%$, and hence the D-instanton effects are unsuppressed. Assuming the axion potential arises from $W_{\rm np}$, the mass and coupling to photons is (see Eq.\eqref{eq:ma_and_ga_expressions})
\begin{equation}
    m_a^2 \sim \frac{\mathcal{A}m_{3/2}M_{\rm GUT}^3}{F_a^2} e^{-S_D}\,,\,\,\,\, g_{a\gamma}\sim \frac{\alpha_{\rm em}}{2\pi F_a}\sim 10^{-19} \text{ GeV}^{-1}\,,
\end{equation}
with $F_a \sim M_{\rm GUT}$ and the D$(-1)$ or D1-instanton action given by the size of the threshold correction, $S_D\sim \delta\alpha_i^{-1}$. 

For threshold corrections smaller than 10\% of the tree-level contribution, we have $m_a\gtrsim \mathcal{O}(1)$ TeV  for gravitino masses as low as $m_{3/2}\gtrsim \mathcal{O}(1)$ eV. Such heavy ALP, if produced in the early Universe via the misalignment or UV freeze-in mechanisms, could cause problems with BBN. This is shown by the black triangle in Figure \ref{fig:ga_vs_ma_plot}. ALPs heavier than $\mathcal{O}(10)$ TeV will decay at earlier times and will not cause any trouble with standard cosmology. This in analogy to the moduli problem in string cosmology~\cite{Banks:1995dp}. 

All in all, when non-universal ALPs obtain their mass from a non-perturbative superpotential we do not expect them to be relevant for low-energy phenomenology, but they may have a big impact in early universe cosmology. Only the QCD axion, which mainly comes from $C_4$, and other $C_4$ axions which could mix with it (if there are other 4-cycles intersecting with the GUT divisor) are relevant for low-energy experiments. We remark, however, that axions that couple to gauge bosons via mixing with the QCD axion will also satisfy \eqref{eq:coupling-to-mass-ratio}.
 The available regions in parameter space for the heavy ALPs are the hatched regions in  Figure~\ref{fig:ga_vs_ma_plot}.

\subsubsection*{ALP mass from K\"ahler potential}
Let us now consider a scenario with small thresholds and all corrections arising only from $\delta K_{np}$ (see Section \ref{sec_Kaehlernonp}). In this case one expects $h^{1,1}_-(X_3)+1$ ALPs with masses near the gravitino mass, $m_{3/2}$, and small couplings to photons: 
\begin{equation}
    m_a^2 \sim \frac{\mathcal{A}_Km_{3/2}^2 M_{\rm GUT}^2 e^{-S_D}}{F_a^2}\,,\,\,\,\, g_{a\gamma}\sim \frac{\alpha_{\rm em}}{2\pi F_a}\sim 10^{-19} \text{ GeV}^{-1}\,,
\end{equation}
with $F_a\sim M_{\rm GUT}$ and $S_D\sim \delta\alpha_i^{-1}$, as discussed in Sections \ref{sec:D-1_instantons} and \ref{sec:D1_instantons}. 

For a light gravitino, $m_a\sim m_{3/2} \sim \mathcal{O}(1-10^3)$ eV, we expect the ALPs (e.g. the black star in Fig.~\ref{fig:ga_vs_ma_plot}) to cause serious cosmological problems, since they tend to overproduce DM by many orders of magnitude. For example, even a single axion with mass $m_a\sim \mathcal{O}$(1) eV and decay constant $F_a\sim M_{\rm GUT}$ overproduces DM by a factor of $\sim 10^{9-10}$ for initial misalignment angles $\theta_i\sim 1$. The abundance of this axion could be \textit{regulated} by tuning its initial misalignment angle to be $\theta_i\lesssim 10^{-5}$. If there exist multiple axions with similar masses and decay constants, the required tuning of the initial misalignment angles is enormous because the tuning grows multiplicatively.\footnote{See \cite{Graham:2018jyp,Reig:2021ipa} for a solution to this problem using a long period of low-scale inflation.}

Similar to the case of axion masses from a superpotential, for a heavier gravitino, $m_a\sim m_{3/2}\gtrsim \mathcal{O}$(10) keV, these heavy axions might be constrained or detectable in astrophysical signals and cosmology through their decay into SM particles. Finally, for gravitino masses above $\mathcal{O}(10-100)$ TeV scale we expect $m_a \gtrsim \mathcal{O}$(10) TeV. These heavy ALPs are expected to decay before BBN and do not cause problems in the early universe cosmology. See the hatched region in Fig.~\ref{fig:ga_vs_ma_plot} for the available regions in parameter space for the heavy ALPs.

\section{Conclusion}\label{sec:conclusion}

In this work, we have shown that the coupling-to-mass ratio $g_{a\gamma}/m_a$ for axions in F-theory GUTs obeys a bound of the type \eqref{eq:coupling-to-mass-ratio}. The main arguments are summarised in the diagram in Fig.~\ref{fig:cartoon_diagram}. Except for the rather hypothetical case that the Pfaffian -- that is, the one-loop fluctuation determinant -- for the relevant stringy instantons is vanishingly small, $\mathcal{A}\ll 10^{-66}$, all axions lie at or below the QCD axion band and the bound \eqref{eq:coupling-to-mass-ratio} holds universally. As illustrated in Fig.~\ref{fig:ga_vs_ma_plot}, the discovery of an ALP above the QCD axion band would falsify this large class of unified quantum gravity theories. The same figure also highlights another generic prediction of F-theory GUTs with approximate gauge coupling unification: the existence of very weakly coupled and comparatively heavy ALPs whose decays may leave observable cosmological signatures. Minimal F-theory GUT models suggest a QCD axion mass of order $m_a^{\rm qcd}\sim 0.5\,\mathrm{neV}$, although a precise determination of this value lies beyond the scope of the present work and is an interesting direction for future investigation.
 
The bound to $g_{a\gamma}/m_a$ \eqref{eq:coupling-to-mass-ratio} originates from a direct relation between gauge coupling unification and ALP masses. Hypercharge flux breaking introduces ALPs from the RR fields $C_0$ and $C_2$ that couple to photons, but not to QCD. The action of the D-instantons that break their shift symmetries -- D$(-1)$-branes for $C_0$ and D1-branes for $c^a$ -- is set by the same threshold corrections that split the SM gauge couplings in the UV. In phenomenologically viable models, where these corrections are small and gauge couplings approximately unify near $M_s$, the instanton actions are correspondingly small, the ALP masses are large, and their ratio $g_{a\gamma}/m_a$ falls orders of magnitude below the QCD axion prediction. The GUT-symmetric axion arising from $C_4$, $c_D$, is protected by the parametrically larger D3-brane action, $S_{D3} = 2\pi/\alpha_{\rm GUT}$, and remains as the QCD axion candidate. The decay constant is expected to be near the string scale ($M_s\gtrsim M_{\rm GUT}$) \cite{Gendler:2023kjt,Fallon:2025lvn} and the mass around $m_a^{\rm qcd} \approx 0.5$~neV. This motivates future searches for the QCD axion in the light regime, such as the DMRadio program~\cite{DMRadio:2022pkf,DMRadio:2022jfv,Benabou:2022qpv,DMRadio:2023igr,Ankel:2026zrv} and superconducting resonant frequency cavities~\cite{Berlin:2019ahk,Giaccone:2022pke}. It will be interesting to pursue more precise computations of the QCD axion mass in F-theory GUTs similar to those presented in \cite{Fallon:2025lvn,Chen:2026bxp} and the case of the weakly curved heterotic string \cite{Benabou:2026jtv}.
 
The variation of the string coupling over the internal manifold, generic in F-theory, only strengthens our results. D-instantons localised in regions where $g_s \sim \mathcal{O}(1)$ -- for example, at matter curves supporting the top quark Yukawa, or along divisors with exceptional gauge symmetry groups -- have unsuppressed actions and generate ALP masses at or above the gravitino mass scale. For GUTs based on $E_6$, $E_7$, or $E_8$, where $g_s$ is frozen at $\mathcal{O}(1)$ throughout the GUT divisor, the non-universal ALPs from $C_0$ and $C_2$ are automatically heavy.
 
Models with large threshold corrections, studied in Section~\ref{sec:large_threshold_and_ga/ma}, might appear to offer an escape to our bound \eqref{eq:coupling-to-mass-ratio}, since large corrections imply large instanton actions and potentially lighter ALPs. However, to reproduce the measured low-energy values of the gauge couplings, such models require incomplete GUT multiplets at intermediate scales. These new states increase the UV gauge coupling and, for example, reduce the D3-brane instanton action. This has two main consequences. Firstly, the UV instanton potential for the would-be QCD axion can dominate over the IR QCD contribution, reintroducing the strong CP problem. Furthermore, achieving a sufficiently large D-instanton action to place an ALP above the QCD band requires internal cycles smaller than the string length, outside the regime of perturbative control in $\alpha^\prime$ corrections. We have verified that these conclusions hold in the most general case, including compactifications with multiple orientifold-odd cycles and fine-tuned cases with $g_s^{a} \ll g_s^{D}$ together with a very light gravitino from gauge mediated SUSY breaking -- a scenario that may be difficult to realise in F-theory GUTs, where $M_s \gtrsim M_{\rm GUT}$.

As discussed in Section \ref{sec:Pfaffian_suppression}, in the extreme case where the Pfaffian takes vanishingly small values, $\mathcal{A}\ll 10^{-66}$, we can also derive upper bounds on $g_{a\gamma}/m_a$. In this case, the contribution to the superpotential and K\"ahler potential from D$(-1)$/D1-branes is negligible, but an irreducible potential for the non-universal ALP arises from its couplings to $SU(2)_w$ and $U(1)_Y$ gauge bosons. We have estimated $g_{a\gamma}/m_a$ by computing the axion potential arising from fluxed D3-instantons. In this case, the axion potential depends on $\alpha_{\rm GUT}$ and the gravitino mass. Figure \ref{fig:ga_vs_ma_plot} contains a benchmark example where $\alpha_{\rm GUT}\sim 1/24$ and $m_{3/2} = 1$ TeV; axions above this line are incompatible with F-theory GUTs even for vanishing Pfaffian factors, $\mathcal{A}$. The coupling-to-mass ratio associated to this line remains orders of magnitude below the minimum required for cosmic birefringence.
 
F-theory GUTs are among the most developed top-down constructions of unified theories, incorporating genuinely non-perturbative strong-coupling effects, a varying dilaton, and topological symmetry breaking mechanisms with no analogue in weakly coupled string theories or in 4d field theory GUTs. That the axion-photon coupling nevertheless delineates a sharp, falsifiable boundary in the $(m_a, g_{a\gamma})$ plane is a non-trivial outcome. It demonstrates that the framework of F-theory GUTs is experimentally testable at current and next-generation axion experiments, from haloscopes looking for axion DM and other DM-independent laboratory searches, to astrophysical observations and CMB polarisation measurements. The most dramatic test is cosmic birefringence: current data contain hints of an axion-induced signal~\cite{Minami:2020odp,Eskilt:2022cff,Diego-Palazuelos:2023mpy,Galaverni:2023zhv,Diego-Palazuelos:2025dmh,Carralot:2026kps} requiring $g_{a\gamma}/m_a \gtrsim 5 \times 10^{15}$~GeV$^{-2}$, deep in the region excluded by our bound \eqref{eq:coupling-to-mass-ratio}. If confirmed in future CMB experiments~\cite{SimonsObservatory:2018koc,2023RScI...94l4502M,LiteBIRD:2022cnt} and uniquely attributed to axions, the signal would rule out F-theory GUTs -- and GUT constructions more generally \cite{Agrawal:2022lsp,Agrawal:2025rbr,Agrawal:2024ejr,Reig:2025dqb}
 -- in the regime of parametric control of the effective theory as a viable UV completion of the Standard Model.

\section*{Acknowledgements}
We thank Josh Benabou, Inaki Garcia-Etxebarria, Markus Dierigl, Damian van de Heisteeg, John March-Russell, Dieter L\"ust, Fernando Marchesano, Luca Martucci, Liam McAllister, Ethan Torres, Alexander Westphal, and Max Wiesner for very useful discussions. We also thank Prateek Agrawal for multiple collaborations in related projects. M.N. would like to acknowledge GRASP Initiative funding provided by Harvard University. M.N. is supported by NSF Award PHY-2310717.

The work of  T.W. is supported in part by Deutsche Forschungsgemeinschaft under Germany’s Excellence Strategy EXC 2121 Quantum Universe 390833306, by Deutsche Forschungsgemeinschaft through a German-Israeli Project Cooperation (DIP) grant “Holography and the Swampland” and by Deutsche Forschungsgemeinschaft through the Collaborative Research Center 1624 “Higher Structures, Moduli Spaces and Integrability.” This article is based in part upon work from COST Action COSMIC WISPers CA21106, supported by COST (European Cooperation in Science and Technology).

\appendix

\section{Corrections to the K\"ahler potential  from D$(-1)$ and D1-instantons}\label{app:NP_corrections_to_K}

In ref.~\cite{Robles-Llana:2006hby} the exact, non-perturbative correction to the K\"ahler potential in the effective action of 4d ${\cal N} = 2$ Type IIB compactifications was obtained by using supersymmetry and the $SL(2, \mathbb Z)$  duality symmetry. It is expressed in the form of a tensor potential, $\chi$, a real function from which we can obtain the K\"ahler potential, $\mathcal{K}=-\log(\chi)$, with
\begin{equation}
    \chi=\chi_{\rm tree}+\chi_{\rm pert}+\chi_{\rm np} \,,
\end{equation}
where $\chi_{\rm np}=\chi_{D(-1)}+\chi_{D1}$ contains contributions from the D$(-1)$-instantons, and fundamental strings and D1-branes. 

In this section we use these results to understand the expected corrections in Calabi-Yau orientifolds. After discussing the corrections in the $\mathcal{N}=2$ theory we will implement the orientifold projection, where in particular $\int_{\Sigma^-}J=0$ for orientifold odd curves $\Sigma^-$. As explained in the main text, it is expected that the corrections that survive the projection do not vanish in the $\mathcal{N}=1$ theory, which is less protected against quantum corrections. The resulting protected part of the action -- that is, corrections calculated from the $\mathcal{N}=2$ theory -- will likely not be the full K\"ahler potential. In particular for $\mathcal{N}=1$ the factorization of K\"ahler and complex structure moduli spaces breaks down and new, genuinely ${\cal N}=1$ quantum effects will correct the tree-level action.  However, the corrections inherited from the 4d ${\cal N}=2$ theory should at least provide a conservative estimate of the non-perturbative corrections that one expects. 

Let us consider first the contribution from D$(-1)$-branes. After Poisson resummation \cite{Robles-Llana:2006hby}, their contribution is given by
\begin{equation}
    \chi_{(-1)} =  \frac{\chi_E}{2(2\pi)^3}\tau_2^{1/2} \left [ 2\zeta(3)\tau_2^{3/2} +\frac{2\pi^2}{3}\tau_2^{-1/2} + 8\pi^2\tau_2^{1/2}\sum_{m\neq 0,n>0} |m/n|e^{2\pi i mn\tau_1}K_1(2\pi|mn|\tau_2)  \right ]  \,.
\end{equation}
The first part comes from worldsheet perturbation theory, the second arises from string corrections at one-loop and the last corresponds to the sum over D$(-1)$-instantons, which we call $\chi_{D(-1)}$. The latter is particularly important for us, given that it is the only correction which depends on the RR axion. Here, $\chi_E$ is the Euler number of the Calabi-Yau, $\tau=\tau_1+i\tau_2=C_0+i/g_s$ is the complex axio-dilaton,\footnote{The axio-dilaton was denoted by the symbol $S$ in the main text, see \eqref{eq:SC2def}. For ease of comparison with \cite{Robles-Llana:2006hby}, we adopt the notation of this reference in this appendix.} and  $K_1(x)$ is a modified Bessel function that decays exponentially for large $x$, $K_1(x)\sim \sqrt{\frac{1}{x}}e^{-x}$. Taking the leading contribution $m=n=1$, for small string coupling, one finds
\begin{equation}
    \chi_{D(-1)}\sim \frac{\tau_2}{2\pi^{3/2}}\chi_E e^{-2\pi\tau_2+2\pi i \tau_1}\,.
\end{equation}

Focusing now on the contributions from $(m,n)$-strings -- that is, bound states of $m$ D1- and $n$ fundamental strings -- one obtains \cite{Robles-Llana:2007bbv}
\begin{align}\label{eq:D1_F1_non-pert_correction}
    &\chi_{1} =- \frac{\tau_2^{1/2}}{(2\pi)^3}\sum _{k_a}n_{k_a}\sum_{m,n} \frac{\tau_2^{3/2}}{|m\tau + n|}(1+2\pi |m\tau + n|k_at^a)e^{-S_{m,n}}\,,\\&
    \text{with }\,\,\, S_{m,n} = 2\pi k_a (|m\tau + n|t^a-imc^a-inb^a)\,.
\end{align}
The coefficients $n_{k_a}$ are the Gopakumar-Vafa invariants that count the rational curves of class $k_a$ in the Calabi-Yau and the indices $m, n$ correspond to sums over D1- and worldsheet instantons, respectively. Here  
$c^a$ is the odd axion from $C_2$, and we define the complex K\"ahler moduli $z^a=b^a+it^a$. 

For simplicity let us focus on the corrections induced by D1-branes, i.e. $m\neq 0$, wrapping a single 2-cycle (see App.~\ref{app:odd_cycles} for $h^{1,1}_-(X_3)>1$), so that we restrict ourselves to $b^a=b\,,c^a=c$; $k^a=k$. Following \cite{Saueressig:2007dr,Robles-Llana:2007bbv} one can perform a Poisson resummation over the index $n$, obtaining
\begin{equation}\label{eq:N=2_D1_non-pert_correction}
    \chi_{D1} = - \frac{\tau_2}{2\pi^2}\sum_{k_a} n_{k_a} \sum_{m\neq 0, n} \frac{|z+n|}{|m|}K_1 (2\pi |m\tau_2||z+n|)e^{2\pi i m (c-\tau_1(b+n))}\,.
\end{equation}
Eq.~\eqref{eq:N=2_D1_non-pert_correction} is a convenient form to express the correction, since it can be more easily connected to the standard D-instanton action, allowing us to estimate the axion induced potential.

For $m=1,n=0$, which corresponds to the correction by a single D1-brane, we obtain a correction
\begin{equation}
  \chi_{D1}^{1,0}   \sim \frac{\tau_2}{2\pi^2} n_k|z| K_1(2\pi \tau_2 |z|)e^{2\pi i (c-\tau_1 b)}+h.c\,.\,,
\end{equation}
which for the orientifold case, where $t=\int_{\Sigma^-} J = 0$ and the argument $2\pi \tau_2|z|=2\pi |b|/g_s\gtrsim
1$, can be re-written as (using the notation in the main text)
\begin{equation}
   \chi_{D1}^{1,0}\sim e^{-2\pi b/g_s}\cos (2\pi (c-C_0 b))\,.
\end{equation}

This result is reassuring because it is the same as what we would naively get by assuming a non-perturbative correction of the type (instanton + anti-instanton)
\begin{equation}
    \delta K_{\rm np}\sim e^{{-}2\pi iG}+e^{2\pi i\bar{G}} = e^{2\pi \text{Im}\,G}\cos (2\pi\, \text{Re}\,G)\,,
\end{equation}
with $G = c-\tau b = (c-C_0b)-ib/g_s$, in agreement with the assumed non-perturbative correction \eqref{eq:non_pert_Kahler_Ga}.

\section{F-theory examples of varying $g_s$}
\label{app:varying_gs}
In this Appendix we give some examples of how $g_s$ varies throughout the compact manifold. We discuss how $g_s$ is sourced and what the expectations are for different GUT gauge groups.

\subsection{Sourcing a dilaton profile}
7-branes are magnetically charged under the RR-axion $C_0$, which combines with the dilaton into the complex axio-dilaton $S = i/{g_s} + C_0$. These objects source a non-trivial profile for $S$ along the direction normal to the divisor $D$ that they wrap,
\begin{equation}\label{eq:dilaton_sourced_7-brane}
    S (z) = S_0 + \frac{1}{2\pi i}\ln (z-z_0) + (\text{regular terms in } z)\,.
\end{equation}
Here $z=z_0$ denotes the position of the 7-brane. As a result of this non-trivial position dependence for the string coupling along the base, $B_3$, the action of the D-instantons can change depending on their location. We will follow the discussion in~\cite{Weigand:2010wm,Weigand:2018rez}.

In generic situations, the string coupling $g_s$ can vary strongly throughout the base $B_3$, taking vanishing values in some places and becoming large in others. For example, when $7$-branes and $O7$-planes intersect, the string coupling vanishes at the position of perturbative $7$-branes and diverges near the position of orientifold O7-planes. For this reason, while eq.~\eqref{eq:dilaton_sourced_7-brane} suggests that the sourced profile (and therefore the string coupling) varies in the directions normal to the divisor $D$, one can also obtain a string coupling $g_s$ that changes along the GUT divisor by locating several non-perturbative objects in the compact space. This means that $S_0=S_0(u_i)$ is a function of the transverse coordinates along the divisor, $u_i$.

The variation of $S$ is modelled as the variation of the complex structure of an elliptic curve transverse to the location of 7-branes. In general the complex structure depends on the base coordinates. This can be described locally by the Jacobi $j$-function,
\begin{equation}
    j(S) = e^{-2\pi i S}+744+196884e^{2\pi i S}+...
\end{equation}
The complex structure $S$ of a Weierstrass elliptic curve can be obtained via the $j$-function 
\begin{equation}
    j(S) = \frac{4(24 f)^3}{\Delta} \,\text{ with} \, \, \, \,\Delta = 4f^3 +27g^2\,.
\end{equation}
In this equation $f,g$ transform as sections of suitable holomorphic  line bundles on $B_3$. The 7-branes sit at the point where the discriminant vanishes $\Delta = 0$.

The different types of 7-branes and associated 4d gauge algebras are encoded in the singularity structure, more precisely the vanishing orders of $(f,g,\Delta)$. If $\Delta$ has a zero of order $N$ while $f$ and $g$ are non-vanishing, this is interpreted as the divisor being wrapped by $N$ coincident perturbative D7-branes, yielding an $SU(N)$ group. In more complicated situations, for other vanishing orders of $(f,g,\Delta)$, one can recover different groups. In particular, and unlike in weakly coupled type IIB, this includes \textit{exceptional} singularities which correspond to exceptional algebras like $E_6,E_7$ and $E_8$ in the worldvolume of stacks of mutually non-local [p,q] 7-branes. $Spin(10)$ and similar GUTs with spinors also become possible. The appearance of exceptional groups and spinor representations is a strong coupling phenomenon not present in the perturbative type II string theories. These examples will be considered in more detail in Section~\ref{sec:specific_GUT_examples}.

\subsection{Weakly varying $g_s$ and Sen's orientifold limit} 
The variation of the dilaton in cases where an $O7$-plane intersects a divisor is well understood within the so-called Sen orientifold limit. This is a well-controlled limit in which an approximately constant and perturbative string coupling $g_s\ll 1$ can be obtained. Let us take the ansatz
\begin{align}\label{eq:sen_orientifold_functions}
    &f = -3h^2+\epsilon \eta \,,\\ &
    g = -2h^3 + \epsilon h \eta -\frac{\epsilon^2}{12}\chi\,,
\end{align}
with $\epsilon$ being a constant and $h,\eta , \chi$ holomorphic functions. For small $\epsilon$, the string coupling is weak everywhere except at $h=0$ where $g_s \rightarrow \infty$. The latter corresponds to the position of the orientifold $O7$-plane. In this case the $j$-function reads $j(S)\simeq \frac{f^3}{\Delta_\epsilon}$, with $\Delta_\epsilon = -9\epsilon^2 h^2 (\eta^2-h\chi)$. From this, we obtain that away from the orientifold the string coupling is
\begin{equation}\label{eq:gs_Sen_limit}
    \frac{2\pi}{g_s}\simeq 4\ln h - 2\ln \epsilon\,.
\end{equation}
The part $\ln h $, where $h$ is the distance to the $O7$-plane, describes the local variation of $g_s$ along the divisor. Far from the $O7$-plane the term $\ln \epsilon$ dominates and $g_s$ can be made approximately constant and perturbatively small, with the concrete value controlled by the constant $\epsilon$.

\subsection{Specific GUT examples}\label{sec:specific_GUT_examples}

With this preparation we now consider common GUT gauge groups in turn and discuss the expected values of $g_s$ for each case. When the variation of the dilaton along the divisor is mild, the results of Section \ref{sec:g_a/m_a_and_unification}, including Fig \ref{fig:ga/ma_vs_action_from_Wnp_new} and \ref{fig:ga/ma_vs_action_from_Knp_new}, hold and the non-universal ALPs are heavy. In other cases, the variation of $g_s$ along the GUT divisor is sizeable and takes $\mathcal{O}(1)$ values in some regions. In these situations the non-universal ALPs are in fact heavier due to the varying $g_s$ effects outlined in Sections  \ref{sec:varying_gs_D(-1)} and \ref{sec:varying_gs_D1}.

\subsubsection{$SU(5)$  with top Yukawa coupling}
%

Even if the GUT group is engineered directly from $SU(5)$, any realistic construction requires non-perturbative physics in some point of the GUT divisor \cite{Donagi:2008ca,Beasley:2008kw, Beasley:2008dc,Donagi:2008kj}. In such models, for example, the top quark Yukawa coupling requires an enhancement of the fiber to $E_6$ over points on the GUT divisor, where $g_s$ is order one. 
 In its neighbourhood, non-perturbative effects will be unsuppressed. 
 In \cite{Cecotti:2009zf} it was even argued that gauge fluxes compatible with a realistic Yukawa coupling structure stabilize D3-branes near these $E_6$ points. Modulo competing effects, this would give a dynamical reason why D$(-1)$-instantons, which like D3-branes are pointlike along the internal space, are expected to localise in this region of small instanton suppression and contribute to the non-perturbative superpotential for $C_0$. However, care  has to be taken because additional, competing background fluxes (possibly in combination with non-perturbative effects) may in principle lead to a net stabilisation effect away from these strongly coupled regions.

\subsubsection{Exceptional GUTs}

Unlike in weakly coupled type II strings, F-theory includes the possibility of  exceptional groups and gauge theories based on $E_6$, $E_7$ and $E_8$ become available. In this section we show that our results also hold in these cases. 

Let us consider the case of $E_8$. Following Tate's algorithm (see \cite{Weigand:2010wm,Weigand:2018rez}), if the order of zeroes is $(f,g,\Delta)=(4,5,10)$ along a divisor, one obtains an $E_8$ singularity leading to an $E_8$ gauge group in the worldvolume of the 7-branes. In this case the complex dilaton is frozen at a constant value $S = e^{2\pi i /3}$ along the GUT divisor. This corresponds to a string coupling $g_s \approx 1.15$. Similar results follow for $E_7$ and $E_6$ where the dilaton takes values $S = e^{2\pi i /4} = i$, and $S = e^{2\pi i /3}$, respectively. These values highlight the non-perturbative behaviour of the exceptional branes. 

In either of the exceptional branes, $E_6$, $E_7$, and $E_8$ we have $g_s \sim \mathcal{O}(1)$. For this reason we expect the action of a D$(-1)$-brane at any point of the GUT divisor or the action of a D1-brane wrapping odd cycles (also within the divisor) to be small, see Eqs.~\eqref{eq:threshold_and_instanton_action} and \eqref{eq:odd_inst_action}. This again implies that the non-universal ALPs obtain large non-perturbative masses and our results hold. For the same reason that the D-instanton actions are small, in these non-perturbative models the non-universal threshold corrections are suppressed due to a large string coupling $g_s$ and the gauge couplings approximately unify near the scale $M_{\rm GUT}$. For recent studies of GUT models based on higher exceptional groups see \cite{Li:2022aek,Li:2023dya}

\subsubsection{$SO(N)$ GUTs with spinor representations}

One way to get $SO(10)$ GUTs or similar groups with spinor representations (otherwise chiral fermions are not possible in $SO(N)$ groups) is to start with $E_6$ and break it down to $SO(10)$ by some symmetry breaking mechanism such as turning on fluxes, Wilson lines or Higgses. In this case the results explained above will apply and the string coupling will be nearly constant and non-perturbative throughout the GUT divisor. 

Another possibility is to start by having a $SO(10)$ GUT gauge group at weak coupling. In this case, which resembles the way that $SU(N)$ groups are obtained for weakly coupled type IIB, one can only obtain spinor representations along a matter curve $C$ when the singularity in the fiber over $C$ is enhanced from $SO(10)$ to $E_6$ or similar non-perturbative groups~\cite{Beasley:2008dc, Beasley:2008kw,Donagi:2008ca, Donagi:2008kj}. In either case, around the curve $C$ the ambient string coupling will be of order $g_s\sim \mathcal{O}(1)$.

\section{The case with $h^{1,1}_-(X_3)>1$}\label{app:odd_cycles}
In this section we consider the case where there exist more than one orientifold odd cycle, $h^{1,1}_-(X_3)=N_->1$, and hence, more than one axion from the RR 2-form field, $c^a$. In this case, the contribution of the odd cycles to the threshold correction is 
\begin{equation}
   \delta \alpha_i^{-1}|_{\rm odd} = \sum_a \delta \alpha_i^{-1}|_{\rm odd}^{a} = \sum_a \delta_i\frac{b^ap_{Ya}}{g_s^{a}}\,.
\end{equation}
Assuming there is no precise cancellation between the different terms, apparent unification in this case requires the sum of corrections, $\delta \alpha_i^{-1}|_{\rm odd}$, to be small when compared to the MSSM prediction $2\pi/\alpha_{\rm GUT}^{\rm mssm}$ (or similar benchmark unified value). 

Let us now show that $N_-=1$ is actually the most conservative case with respect to the lightness of the ALP $c^a$ (that is the case which allows the lightest $c^a$). To this end, for simplicity, let us assume that $p_{Ya}=1$ and $b^a\sim 1$, for $a=1,...,N_-$. For simplicity we also assume that the average string couplings over the different odd cycles are comparable. Note that if there is a hierarchy between the axion VEVs $b^a$ divided by string couplings $g_s^{a}$, one should focus on the contribution that maximizes $\delta \alpha_i^{-1}|_{\rm odd}^{a}$. Such situation would resemble a case with $N_-=1$. All in all, in this simplified situation we obtain that the correction to the gauge coupling scales with the number of odd cycles as
\begin{equation}
    \delta \alpha_i^{-1}|_{\rm odd} \sim \frac{N_-}{g_s^{a}}\,.
\end{equation}
This allows us to write the action of a D1-brane wrapping the $a$-th odd cycle as
\begin{equation}
    S_{D1}^a =\frac{2\pi b^a}{g_s^{a}} = \frac{2\pi |\delta\alpha_i|_{\rm odd}^{-1}}{N_-}\,.
\end{equation}

The lightness of the ALPs $c^a$ is maximised for the largest D1-brane action, which for a fixed threshold correction $\delta \alpha_i^{-1}|_{\rm odd} $, occurs for $N_-=1$. This indicates that the case considered in Section \ref{sec:D1_instantons} is conservative, as it is the case where one expects $c^a$ to obtain the smallest possible mass from D1-branes wrapping odd cycles.

\bibliographystyle{JHEP}

\providecommand{\href}[2]{#2}\begingroup\raggedright\endgroup

\end{document}